\documentclass{aa}

\usepackage{graphicx}
\usepackage{txfonts}
\usepackage{color}
\newcommand{\cm}{\mathrm{cm}}
\newcommand{\g}{\mathrm{g}}
\newcommand{\yr}{\mathrm{year}}
\newcommand{\yrs}{\mathrm{years}}

\newcommand{\AU}{{\rm AU}}
\newcommand{\Mearth}{M_{\mathrm{\oplus}}}
\newcommand{\Mjupiter}{M_{\mathrm{J}}}
\newcommand{\Msun}{M_{\mathrm{\odot}}}
\newcommand{\Gconst}{G}

\newcommand{\deriv}[2]{\frac{\mathrm{d} #1}{\mathrm{d} #2}}
\newcommand{\dderiv}[2]{\frac{\mathrm{d^2} #1}{\mathrm{d} {#2}^2}}

\newcommand{\Mstar}{M_{ {\rm s} }}
\newcommand{\Mplanet}{M_{ {\rm p} }}
\newcommand{\mplts}{m_{ {\rm pl} }}
\newcommand{\Rplts}{R_{ {\rm pl} }}
\newcommand{\axiplts}{a}
\newcommand{\axipltsint}{a_{ {\rm 0} }}

\newcommand{\axiplanet}{a_{ {\rm p} }}
\newcommand{\axiplanetint}{a_{\rm p,int}}
\newcommand{\axiplanetfnl}{a_{\rm p,fnl}}
\newcommand{\rhill}{h_{\rm }}
\newcommand{\fgas}{f_{\rm gas}}
\newcommand{\fgrav}[2]{f_{\rm grav}({\it #1},{\it #2})}
\newcommand{\fmig}{f_{\rm tid}}

\newcommand{\cd}{C_{\rm d}}
\newcommand{\rhogas}{\rho_{\rm gas}}
\newcommand{\vplgs}{u}
\newcommand{\vgas}{v_{\rm gas}}
\newcommand{\Reyn}{\mathcal{R}}
\newcommand{\Mach}{\mathcal{M}}
\newcommand{\vplts}{v_{\rm pl}}
\newcommand{\vplanet}{v_{\rm p}}

\newcommand{\Sigmagas}{\Sigma_{\rm gas}}    
\newcommand{\Sigmasolid}{\Sigma_{\rm solid}}

\newcommand{\Sigmazero}{\Sigma_{\rm ss}}
\newcommand{\csound}{c_{\rm s}}
\newcommand{\hscale}{h_{\rm s}}
\newcommand{\OmegaK}{\Omega_{\rm K}}
\newcommand{\vK}{v_{\rm K}}
\newcommand{\alphaCSD}{\alpha_{\rm v}}
\newcommand{\Mdisczero}{M_{\rm disc,0}}

\newcommand{\tss}{\tilde{t}}
\newcommand{\tauss}{\tau_{\rm s}}
\newcommand{\tzero}{t_{\rm 0}}

\newcommand{\Rone}{R_{\rm 1}}
\newcommand{\Rtwo}{R_{\rm 2}}

\newcommand{\etadisc}{\eta}
\newcommand{\Pgas}{P_{\rm gas}}

\newcommand{\Zstar}{Z_{\rm s}}

\newcommand{\fdisc}{f_{\rm disc}}
\newcommand{\fsolid}{f_{\rm solid}}
\newcommand{\Tice}{T_{\rm ice}}
\newcommand{\nbody}{N_{\rm sp}}
\newcommand{\axipltsin}{a_{\rm in}}
\newcommand{\axipltsout}{a_{\rm out}}

\newcommand{\Msp}{m_{\rm sp}}

\newcommand{\rhoplanet}{\rho_{\rm p}}

\newcommand{\rplts}{r_{\rm pl}}
\newcommand{\rplanet}{r_{\rm p}}
\newcommand{\rstar}{r_{\rm s}}
\newcommand{\Ejacobi}{E_{\rm J}}
\newcommand{\Ujacobi}{U_{\rm J}}

\newcommand{\Mcaptot}{M_{\rm cap}^{\rm total}}

\newcommand{\lmfp}{l_{\rm m}}
\newcommand{\Ocrr}{\omega_{\rm crr}}
\newcommand{\tauaero}{\tau_{\rm aero}}

\newcommand{\tautidezero}{\tau_{\rm tid,0}}
\newcommand{\tautideaxi}{\tau_{\rm tid,a}}

\begin{document} 

    \title{The origin of the high metallicity of close-in giant exoplanets} 
    \subtitle{Combined effect of the resonant and aerodynamic shepherding}

    \author{
            Sho Shibata
            \inst{1}
            \thanks{E-mail: s.shibata@eps.s.u-tokyo.ac.jp}
        \and
            Ravit Helled
            \inst{2}
        \and
            Masahiro Ikoma
        \inst{1,3}
        }
        
    \institute{
            Department of Earth and Planetary Science, Graduate School of Science, The University of Tokyo, 7-3-1 Hongo, Bunkyo-ku, Tokyo 113-0033, Japan
        \and
            Institute for Computational Science, 
            Center for Theoretical Astrophysics \& Cosmology, 
            University of Zurich
            Winterthurerstr. 190
            CH-8057 Zurich
            Switzerland
        \and
            Research Center for the Early Universe (RESCEU),
            Graduate School of Science, The University of Tokyo, 7-3-1 Hongo,
            Bunkyo-ku, Tokyo 113-0033, Japan
    }

    \date{Received September XX, YYYY; accepted September XX, YYYY}

    \abstract
    {
    Recent studies suggest that many giant exoplanets are highly enriched with heavy elements compared to their host star and contain several tens of   Earth masses or more of heavy elements.
    Such enrichment is considered to have been brought by accretion of planetesimals in late formation stages.
    Previous dynamical simulations, however, show that planets are unable to collect so much heavy elements through {\it in situ} planetesimal accretion.
    }
    {
    We investigate whether a giant planet migrating inward can capture planetesimals efficiently to significantly increase its metallicity.  
    }
    {
    We performed orbital integrations of a migrating giant planet and planetesimals in a protoplanetary gas disc to infer the planetesimal mass that is accreted by the planet.
    }
    {
    We find that the two shepherding processes of mean motion resonances trapping and aerodynamic gas drag  inhibit planetesimal capture of a migrating planet. However, the amplified libration allows the highly-excited planetesimals in the resonances to escape from the resonance trap and be accreted by the planet.
    Consequently, we show that a migrating giant planet captures planetesimals with total mass of several tens of Earth masses, if the planet forms at a few tens of AU in a relatively massive disc.
    We also find that planetesimal capture occurs efficiently in a limited range of semi-major axis, and that the total captured planetesimal mass increases with increasing  migration distances.  
    Our results have important implications for understanding the relation between giant planet metallicity and mass, as we suggest that it reflects the formation location of the planet, or more precisely, the location where runaway gas accretion occurred.   
    We also suggest the observed metal-rich close-in Jupiters migrated to their present locations from afar, where they formed. 
    }
    {}    


    \maketitle
%

\section{Introduction}
\label{sec:Introduction}
    Many of the detected close-in giant exoplanets are found to be highly enriched with heavy elements compared to stellar composition, with several of the planets containing several hundreds Earth-mass of heavy elements \citep{Guillot+2006,Miller+2011,Thorngren+2016}. The origin of this enrichment, however, is poorly understood. 
    One might naively think that giant planets formed by core accretion have metallicities (or core-to-envelope mass ratios) higher than those of the host stars.
    This is not necessarily true, since the mass of a core formed before the onset of the runaway gas accretion is at most $\sim$ 20-30 Earth masses \citep[e.g.,][]{Pollack+1996,Ikoma+2006b,Lambrechts+2014}.
    Further enrichment of the planet during runaway gas accretion is difficult; indeed, theoretical studies show that only several Earth masses of heavy elements can be captured {\it in situ} during the gas accretion phase, when a minimum-mass solar nebula is assumed \citep{Zhou+2007,Shiraishi+2008,Shibata+2019}.
    If the solid mass is several times higher than the heavy-element mass of the minimum-mass solar nebula, the enrichments of Jupiter and Saturn could be explained by such an {\it in situ} accretion during the gas accretion stages \citep{Shibata+2019}. 
    However, the accreted masses are found to be significantly smaller than the inferred several tens (or more) of Earth masses of the detected warm-Jupiters \citep{Thorngren+2016}.
    In addition, \citet{Thorngren+2016} found a relation between the heavy-element mass and planet mass that has to be explained.

    Enrichment of gas giants during their formation process has been recently investigated by several groups.
    \citet{Bitsch+2019a} studied formation of planetary systems considering pebble accretion and \citet{Booth+2017} investigated the enrichment of the planetary envelop considering the composition evolution of disc gas due to pebbles.
    Their results suggest that pebble accretion is insufficient to explain the inferred amount of heavy elements in close-in giant planets.
    Indeed, the asteroid belt population in our solar system suggests that large planetesimals do form, and could be available to capture during the planetary growth \citep{Morbidelli+2009,Johansen+2015}. In addition, a recent formation model of Jupiter suggests that planetesimal accretion is required to explain the separation of solids in the early solar system \citep{Alibert18}. 
    
    Using empirical formulae for {\it in situ} accretion rates of planetesimals and pebbles available in the literature, \citet{Hasegawa+2018} investigated the metallicity enhancement of close-in planets and found that the metallicity increases with the planetary mass. However, they have not studied the absolute amount of heavy elements in the planet, which remains unknown.
    In this study, we focus on the actual captured mass of planetesimals, and its dependence on various model assumptions. 
    In particular, we investigate the effects of planetary migration on the capture of planetesimals, since most of the giant exoplanets that are found to be enriched have short orbits and therefore are expected to experience significant migration after their formation.

    During its orbital migration, a gas giant planet encounters many planetesimals and captures some of them.
    \citet{Tanaka+1999} performed $N$-body simulations of the dynamics of planetesimals and a migrating protoplanet.
    Jupiter's enrichment with heavy elements during its formation when migration is included was presented by \citet{Alibert+2005}. 
    Both studies neglected the effects of mean motion resonances, which should be considered since they significantly affect the planetesimals orbital evolution  during planetary migration \citep{Batygin+2015}.
    Currently, the efficiency of planetesimal accretion during planetary migration remains unknown.

    The main goal of this study is to investigate the basic physics of planetesimal capture by a migrating gas giant planet, and explore whether this process can explain the inferred enrichments of close-in giant exoplanets.
    Our paper is organised as follows. 
    In Sec.~\ref{sec:Method} we describe the basic model and settings used in this study.
    We construct a reference model, which is explained in Sec.~\ref{sec:Method_Settings}.
    In Sec.~\ref{sec:Result1} we investigate the capture process using the reference model.
    The basic physics of planetesimal capture by a migrating planet is analysed in detail.
    In Sec.~\ref{sec:Result2}, we perform parameter studies and investigate the sensitivity of the inferred captured mass to the assumed parameters.  
    We compare our results with observations in Sec.~\ref{sec:Dis_Compare} and discuss the limitations of our model in Sec.~\ref{sec:Dis_Mod_Eff}.
    A summary of the study and its conclusions are presented in Sec.~\ref{sec:Summary}.

\section{Method and Model}
\label{sec:Method}
    In this study we assume the following situation: 
    A giant planet was formed after gas accretion stopped and the planet is no longer growing in mass. 
    The planet then opens a gap and migrates radially inward in the type-II mode from a given semi-major axis in a circumstellar disc. 
    Initially there are many single-sized planetesimals interior to the planet's orbit.  
    The migrating planet then encounters these planetesimals and captures some of them. 
    Planetesimals are represented by test particles and, therefore are affected only by the gravitational forces from the central star and planet, and the drag force by the disc gas.
    The dynamical integration for these bodies is performed using the numerical simulation  developed in \citet{Shibata+2019}, where the detailed scheme and benchmark results are described.

    \subsection{Forces Exerted on Planetesimals and Planet}\label{sec:Method_Basic_Equations}
        The equation of motion is given by
        \begin{align}\label{eq:equation_of_motion}
          \dderiv{ {\bf r}_{\rm i} }{t}    &= \sum_{i\neq j} {\bf \fgrav{i}{j}} + {\bf \fgas} +{\bf \fmig}, 
        \end{align}
        where $t$ is the time, ${\bf r}_i$ is the position vector relative to the initial (i.e., $t = 0$) mass centre of the star-planet-planetesimals system, ${\bf \fgrav{i}{j}}$ is the mutual gravity between particles $i$ and $j$ given by
        \begin{align}
            {\bf \fgrav{i}{j}} &= - \Gconst \frac{M_{\rm j}}{{r_{i,j}}^3} {\bf r}_{i,j}
        \end{align}
        with $\mathbf{r}_{i,j}$ being the position vector of particle $i$ relative to particle $j$ ($r_{i,j} \equiv |{\bf r}_{i,j}|$),
        $M_{\rm j}$ is the mass of particle $j$, and $\Gconst$ is the gravitational constant,
        ${\bf \fgas}$ is the aerodynamic gas drag, and ${\bf \fmig}$ is the gravitational tidal drag from the circumstellar-disc gas. 
        The central star, planet, and planetesimals are denoted by the subscripts $i$ (or $j$) = 1, 2, and $\geq$~3, respectively.
        The planetesimals are treated as test particles; therefore $\fgrav{i}{j}=0$ in Eq.~(\ref{eq:equation_of_motion}) for $j \geq$~3. 
        Also, given the range of the planetesimals mass ($\sim10^{16}$-$10^{22} \g$) and planet ($\sim10^{30} \g$), we assume $\fmig=0$ for the former and $\fgas=0$ for the latter.
        The central star is not affected by $\fgas$ and $\fmig$. 

\subsubsection{Aerodynamic Gas Drag}\label{sec:Method_AeroDrag}
The aerodynamic gas drag force is given by \citep{Adachi+1976}
\begin{align}\label{eq:reduced_gas_drag}
  {\bf \fgas} = -\frac{\bf \vplgs}{\tauaero} = - \frac{1}{2 \mplts} \cd \pi {\Rplts}^2 \rhogas \vplgs {\bf \vplgs}.
\end{align}
Here ${\bf \vplgs}$ is the velocity relative to the ambient gas ($\vplgs = |{\bf \vplgs}|$), $\tauaero$ is the damping timescale of aerodynamic gas drag, $\mplts$ is the planetesimal's mass, $\cd$ is the non-dimensional drag coefficient (see Appendix~\ref{App_Cd}), $\rhogas$ is the gas density, and $\Rplts$ is the planetesimal's radius.
The velocity and density of the ambient disc gas are calculated from the circumstellar disc model (see Section~\ref{sec:Method_Disc}).

\subsubsection{Gravitational Tidal Drag}\label{sec:Method_TidalDrag}
We also consider the effect of the type-II migration of the gas giant planet, using the following form of the tidal drag force,
\begin{align}\label{eq:tidal_drag}
  {\bf \fmig} = - \frac{\bf \vplanet}{2 \tautideaxi},
\end{align}
where ${\bf \vplanet}$ and $\tautideaxi$ are the planet's velocity and the damping timescale of the semi-major axis, respectively. 
In this formula, the planet's eccentricity is assumed to be negligibly small.
As for $\tautideaxi$, we consider type II migration, with the dependence on the planet's semi-major axis $\axiplanet$ being: 
\begin{align}\label{eq:migration_timescale2}
    \tautideaxi = \tautidezero \left(  \frac{\axiplanet}{1 \AU} \right)^{1/2},
\end{align}
where the constant $\tautidezero$ is set as a free parameter.
Equation 5 does not depend on the planet's mass and the disc properties as shown by \citep{Ida+2004a}.
However, this simplification allows a systematic investigation of the effect of planet migration on the efficiency of planetesimal capture and the global enrichment.


    \subsection{Disc Model}\label{sec:Method_Disc}
        We adopt the so-called self-similar solution for the surface density $\Sigmazero$, which is expressed as \citep{Lynden-Bell+1974} 
        \begin{align}
            \Sigmazero   &= \frac{\Mdisczero}{2 \pi {R_{\rm m}}^2} \left(\frac{r}{R_{\rm m}}\right)^{-1} {\tss}^{-3/2} \exp \left( - \frac{r}{\tss R_{\rm m}} \right), \label{eq:disc_surface_density2}\\
            \tss        &= \frac{t_{\rm 0}}{\tauss} + 1, \label{eq:disc_surface_density2_time}
        \end{align}
        where $\Mdisczero$ is the initial total disc mass,
        $R_{\rm m}$ is the radial scaling length of circumstellar disc,
        $\tzero$ is the time when the planetary migration begins, and 
        $\tauss$ is the characteristic time of viscous evolution defined as $\tauss={R_{\rm m}}^2/3 \nu_{\rm m}$ with $\nu_{\rm m}$ being the viscosity at $r=R_{\rm m}$. 
        Equation~(\ref{eq:disc_surface_density2}) is derived under the assumption of the Keplerian-rotating disc with the temperature profile of $T \propto r^{-1/2}$ and the $\alpha$-prescription for turbulent viscosity from \citet{Shakura+1973}.
        It should be noted that $\tss$ remains fixed during the simulation, which means that $\Sigmazero$ depends solely on $t_{\rm 0}$, the time at which the planet starts migrating.  
        This simplification allows a systematic investigation of the planetesimal accretion efficiency. 

\begin{figure}
            \begin{center}
                \includegraphics[width=80mm]{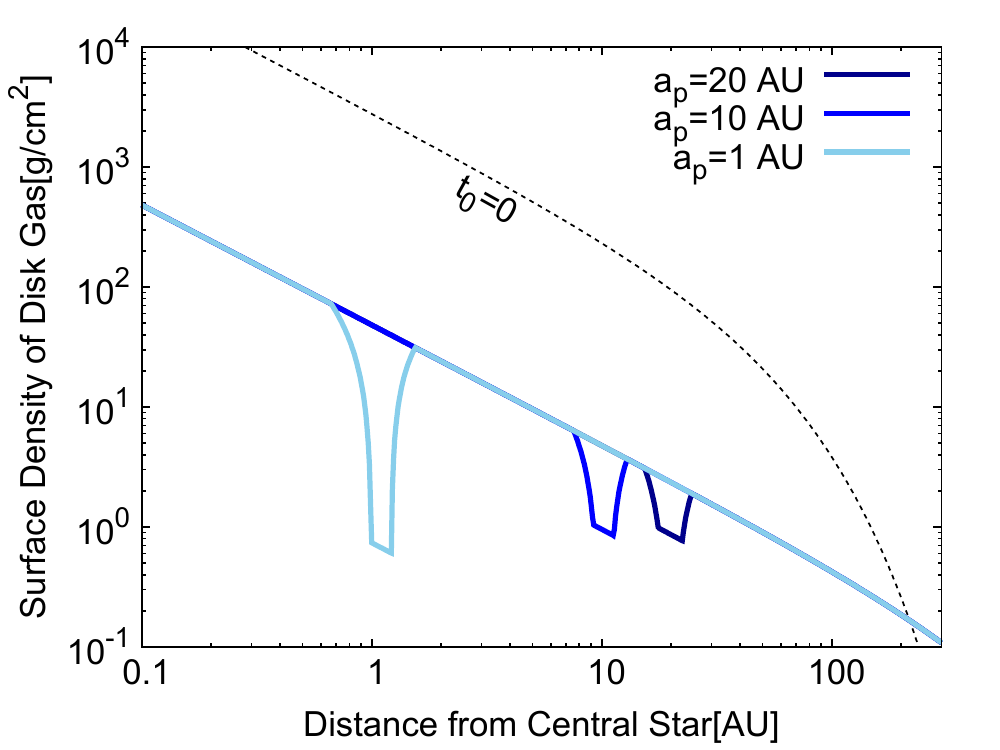}
                \caption{
                    An example of the disc's surface density during planetary migration with $t_{\rm 0}=3\times10^6\yr$.
                    The disc's surface density is shown as a function of distance from the central star. 
                    The solid lines show the radial profiles with gap opening for different locations of the planet as indicated in the legend.
                    The black-dotted line shows the surface density of disc gas for $t_{\rm 0}=0$.
                }
                \label{fig:Surface_Density_Profile}
            \end{center}
        \end{figure}

        The gap's structure is modeled in the same way as \citet{Shibata+2019}; namely, we use the empirical formula for the radial density profile in the gap derived from the two-dimensional hydrodynamical simulations of \citet{Kanagawa+2017}. 
        Combining the surface density profiles of the circumstellar disc and the gap, we express the gas surface density $\Sigma_{\rm gas}$ as
        \begin{align}\label{eq:disc_surface_density_wit_gap}
            \Sigma_{\rm gas} (r) = f_{\rm gap} (r) \, \Sigma_{\rm ss} (r),
        \end{align}
        where $f_{\rm gap}$ is the window function representing the effect of gap opening (see Appendix~\ref{App_disk}).
        Examples of the radial profiles of $\Sigma_{\rm gas}$ are shown in Fig.~\ref{fig:Surface_Density_Profile}.
        Also, the circumstellar disc being assumed to be vertically isothermal, the gas density $\rhogas$ is expressed as
        \begin{align}\label{eq:gas_volume}
            \rhogas = \frac{\Sigmagas}{\sqrt{2 \pi} \hscale} \exp \left( -\frac{z^2}{2 {\hscale}^2} \right),
        \end{align}
        where $z$ is the height from the disc mid-plane and $\hscale$ is the disc's scale height.

        The gas in the circumstellar disc rotates with a sub-Keplerian velocity $\vK$ because of pressure gradient; namely
        \begin{align}\label{eq:sub-Kepler}
	        \vgas = \vK \left( 1-\etadisc \right)
        \end{align}
        with $\etadisc$ defined as
        \begin{align}
	        \etadisc \equiv	\mathbf{-} \frac{1}{2} \left( \frac{\hscale}{r} \right)^2 \deriv{ \ln \Pgas}{\ln r}, 
        \end{align}
        where $\Pgas$ is the gas pressure.
        For deriving the above equation, we assume $\etadisc \ll 1$ and use the ideal-gas relation for isothermal sound speed, i.e., $\csound^2$ = $\Pgas / \rhogas$.

    \subsection{Treatment of planetesimals}
        We assume that the evolution of planetesimal disc is decoupled from that of gas disc and the initial surface density of solids $\Sigmasolid$ is related to the surface density of gas by
        \begin{align}\label{eq:surface_density_solid}
            \Sigma_{\rm solid} (r, t=0) = \fsolid (r) \, \Zstar \, \Sigmazero(r,t_{\rm 0}=0),
        \end{align}
        where $\Zstar$ is the solid-to-gas ratio, which is assumed equal to the metallicity of the central star, and $\fsolid$ is a factor regarding the composition of planetesimals; in this study, we assume the planetesimals are purely rocky and ice mixed with rock interior and exterior to the snowline, respectively, and $f_{\rm solid}$ is given as \citep{Hayashi1981}
        \begin{align}\label{eq:fsolid}
            \fsolid (r) =\left\{
            \begin{array}{ll}
                0.24 & \text{for $T(r) > \Tice$}, \\
                1.0  & \text{for $T(r) < \Tice$},
            \end{array}
            \right.
        \end{align}
        where $\Tice$ is the sublimation temperature of water ice, 170~K.
        The temperature of the circumstellar disc $T$ is given by: 
        \begin{align}
            T=T_{\rm m} \left(\frac{r}{R_{\rm m}} \right)^{-0.5},
        \end{align}
        where $T_{\rm m}$ is the disc gas' temperature at $r=R_{\rm m}$.
        The planetesimal's mass $\mplts$ is calculated as $4\pi\rho_{\rm pl}{\Rplts}^3/3$, where $\rho_{\rm pl}$ is the planetesimal's mean density.

        We follow the orbital motion of super-particles, each of which contain several equal-size planetesimals.
        The super-particles are distributed in a given radial region uniformly, where the inner and outer edges are denoted by $\axipltsin$ and $\axipltsout$, respectively. 
        The surface number density of super-particles is  $\propto r^{-1}$. 
        The mass per super-particle $\Msp$ is given by
        \begin{align}
            \Msp(\axipltsint) = \frac{1}{\nbody}\fdisc~\fsolid(\axipltsint)~\Zstar~\Mstar  ~\frac{\axipltsout-\axipltsin}{R_{\rm m}} \exp\left(-\frac{\axipltsint}{R_{\rm m}}\right),
        \end{align}
        where $\axipltsint$ is the initial semi-major axis of the super-particle, $\nbody$ is the number of super-particles used in a given simulation, and $\fdisc$ is the mass ratio of the initial circumstellar disc to the central star.

        During the orbital integration, we judge that a super-particle has been captured by the planet once (i) the super-particle enters the planet's envelope or (ii) its Jacobi energy becomes negative in the Hill sphere.
        The planet's radius $R_{\rm p}$ is calculated as
        \begin{align}\label{eq:Rplanet_cap}
            R_{\rm p} = \left( \frac{3 M_{\rm p}}{4 \pi \rho_{\rm p}} \right)^{1/3},
        \end{align}
        where $\rho_{\rm p}$ is the planet's mean density, and we set $\rho_{\rm p}=0.125~\g~\cm^{-3}$.
        The Jacobi energy is defined as \citep[e.g.][]{Murray+1999}
        \begin{align}
            \Ejacobi &\equiv    \frac{1}{2} {\vplts^{\prime}}^2 + \Ujacobi, \label{eq:Jacobi_Energy} \\
            \Ujacobi &=         -\frac{1}{2} {\OmegaK}^2 \left( {{x}^{\prime}}^2 + {{y}^{\prime}}^2 \right) 
                                -\Gconst \frac{\Mstar}{\left|{\bf \rplts} - {\bf \rstar}\right|} - \Gconst \frac{\Mplanet}{\left|{\bf \rplts} - {\bf \rplanet}\right|} + U_0 \label{eq:Jacobi_Potential},
        \end{align}
        where ($x^\prime$, $y^\prime$) and $\vplts^\prime$ are, respectively, the position and velocity of the planetesimal in the coordinate system co-rotating with the planet.
        The constant $U_0$ is set such that $\Ujacobi$ vanishes at the Lagrange L$_2$ point.
        With orbital elements, the
        Jacobi energy of a planetesimal is approximately expressed as \citep{Hayashi+1977}
        \begin{align}\label{eq:Ejacobi}
            E_{\rm J} = \frac{G M_{\rm s}}{a_{\rm p}} \left\{ -\frac{a_{\rm p}}{2 a} - \sqrt{\frac{a}{a_{\rm p}} \left(1-e^2\right)} \cos {i} + \frac{3}{2} + \frac{9}{2} h^2 +O(h^3) \right\},
        \end{align}
        where $a$, $e$ and $i$ are the planetesimal's semi-major axis, eccentricity and inclination and $h$ is the reduced Hill radius defined as
        \begin{align}\label{eq:reduced_Hill_radius}
            h = \left( \frac{M_{\rm p}}{3 M_{\rm s}} \right)^{1/3}.
        \end{align}
        The region of $E_{\rm J} > 0$ corresponds the so-called feeding zone, inside which planetesimals can enter the planet's Hill sphere. 

    \subsection{Model Settings}\label{sec:Method_Settings}
        In Sec.~\ref{sec:Result1}, we investigate the basic physics of the process of planetesimal capture by the migrating gas giant planet.
        The planet is initially located at a semi-major axis $a_{\rm p,int}$ and migrates inward as determined by Eq.~(\ref{eq:tidal_drag}).
        The calculation is artificially stopped once the planet reaches the orbit of $a_{\rm p,fnl}$. 
        The planetesimals are initially distributed outside the initial feeding zone of the planet, namely $a_{\rm p,fnl} (1-2\sqrt{3} \rhill) < a < a_{\rm p,int} (1-2\sqrt{3} \rhill)$.
        The choices of the parameter values for the reference model are summarised in Table~\ref{tb:settings}.
        In Sec.~\ref{sec:Result2}, we perform a parameter study for different values of  $\tautidezero$, $\Rplts$, $\axiplanetint$ and $\Mplanet$ and investigate their effect on the captured heavy-element mass.

        Again, in this study, to focus on the effect of planetary migration on the planetesimal capture process, we do not consider planetary growth nor disc evolution. 
        While our setup is simplified, it allows us to identify the parameters that strongly influence the planetesimals orbital evolution and the efficiency of planetesimal capture. This is discussed in detail in Sec.~\ref{sec:Discussion}.

        \begin{table*}
	        \centering
	        \begin{tabular}{ llllll } 
		    \hline
		    Parameters used in reference model &&&&&\\
		    \hline
		        $\Mstar$ 		& Mass of central star	                        & $1.0$				                        & $\Msun$ 	\\
		        $\Zstar$ 		& Metalicity of central star	                & $0.014$				                    & - \\
		        $\Mplanet$ 		& Mass of planet		                        & $1.0$		                                & $\Mjupiter$  \\
                $\axiplanetint$ & Initial semi-major axis of planet             & 20                                & $\AU$ \\
                $\axiplanetfnl$ & Final semi-major axis of planet               & $0.5$                             & $\AU$ \\
		        $\rhoplanet$    & Mean density of planet                        & $0.125$                           & $\g~\cm^{-3}$ \\
		        $\tautidezero$  & Scaling factor of migration timescale	        & $1.0 \times 10^5$                 & $\yr$ \\
                $R_{\rm m}$ 	& Initial size of circumstellar disc            & $50$		                        & $\AU$  \\
		        $T_{\rm m}$ 	& Temperature at $r$ = $R_{\rm m}$              & $40$		                        & K  \\
                $\alphaCSD$     & Viscosity parameter of disc gas               & $1.0 \times 10^{-2}$              & - \\
                $\fdisc$        & Initial disc mass relative to central star    & $0.1$                             & - \\ 
                $\tzero$        & Time of the onset of planetary migration      & $3.0\times10^{6}$       & $\yr$ \\
		        $\Rplts$		& Radius of planetesimal			            & $1.0 \times 10^6$	                & $\cm$		    \\
		        $\rho_{\rm pl}$ & Mean density of planetesimal			        & $2.0$	                            & $\g~\cm^{-3}$		    \\
		        $N_{\rm sp}$    & Initial number of super-particles             & 10 000                  & - \\
		    \hline
	        \end{tabular}
	        \caption{
	            Parameters used in the reference model.
            }
            \label{tb:settings}
        \end{table*}

\section{Reference Model Results}
\label{sec:Result1}
    \subsection{Dynamics of Planetesimals around a Migrating Planet}
    \label{sec:Result1_1}
        \begin{figure}
            \begin{center}
            \includegraphics[width=75mm]{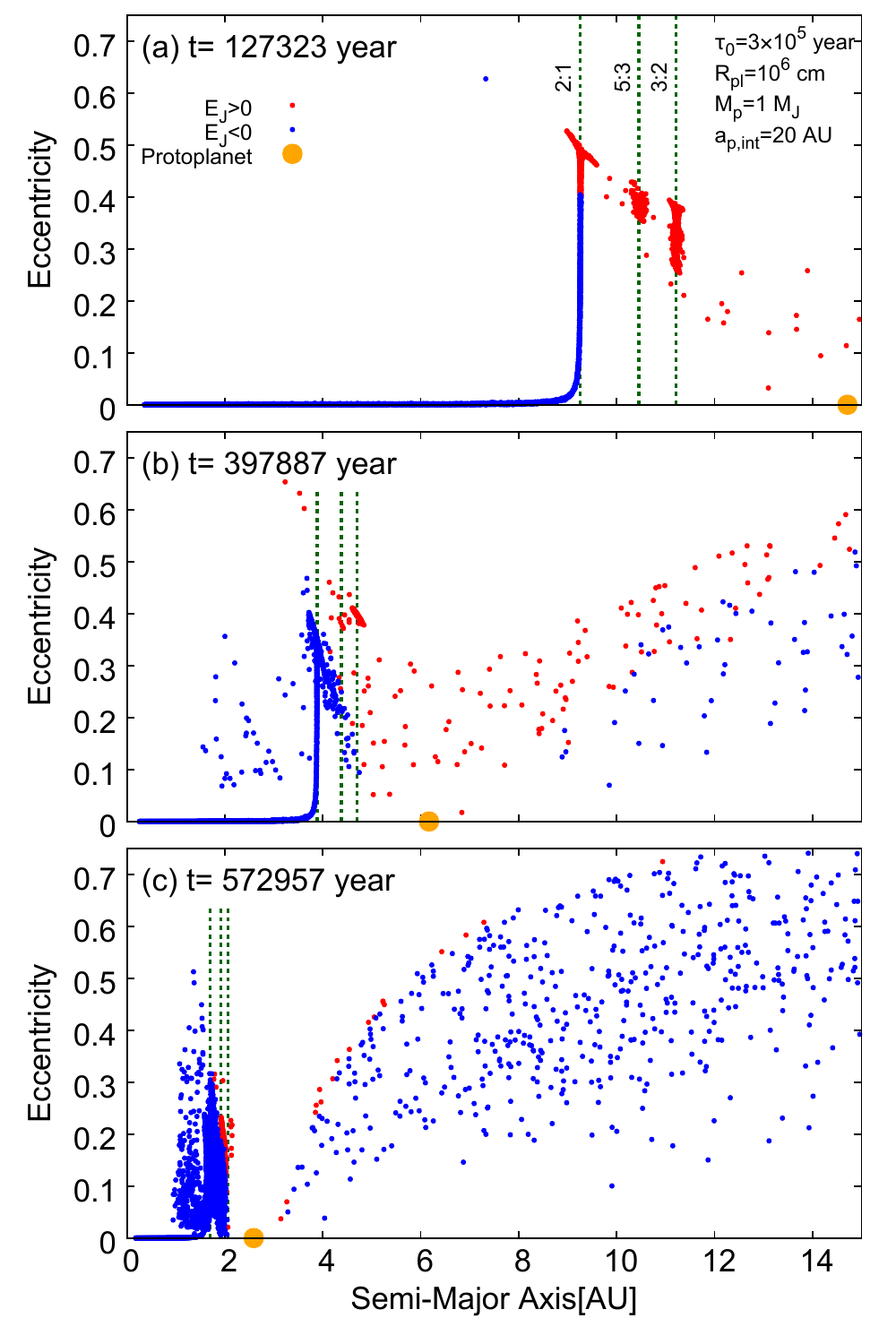}
            \caption{
                Snapshots of the orbital evolution of planetesimals for (a) $t$ = 127,323~yr, (b) 397,887~yr, and (c) 572,957~yr in the reference case (see Table~\ref{tb:settings} for the setting). 
                The horizontal and vertical axes are the semi-major axis and eccentricity, respectively.
                The red and blue circles indicate the planetesimals with positive and negative Jacobi energy, respectively; the orange circle represents the migrating planet.
                The green dotted lines indicate the positions of 2:1, 5:3, and 3:2 mean motion resonances with the planet.
            }
            \label{fig:Result_Snapshots}
            \end{center}
        \end{figure}

        \begin{figure}
            \begin{center}
            \includegraphics[width=80mm]{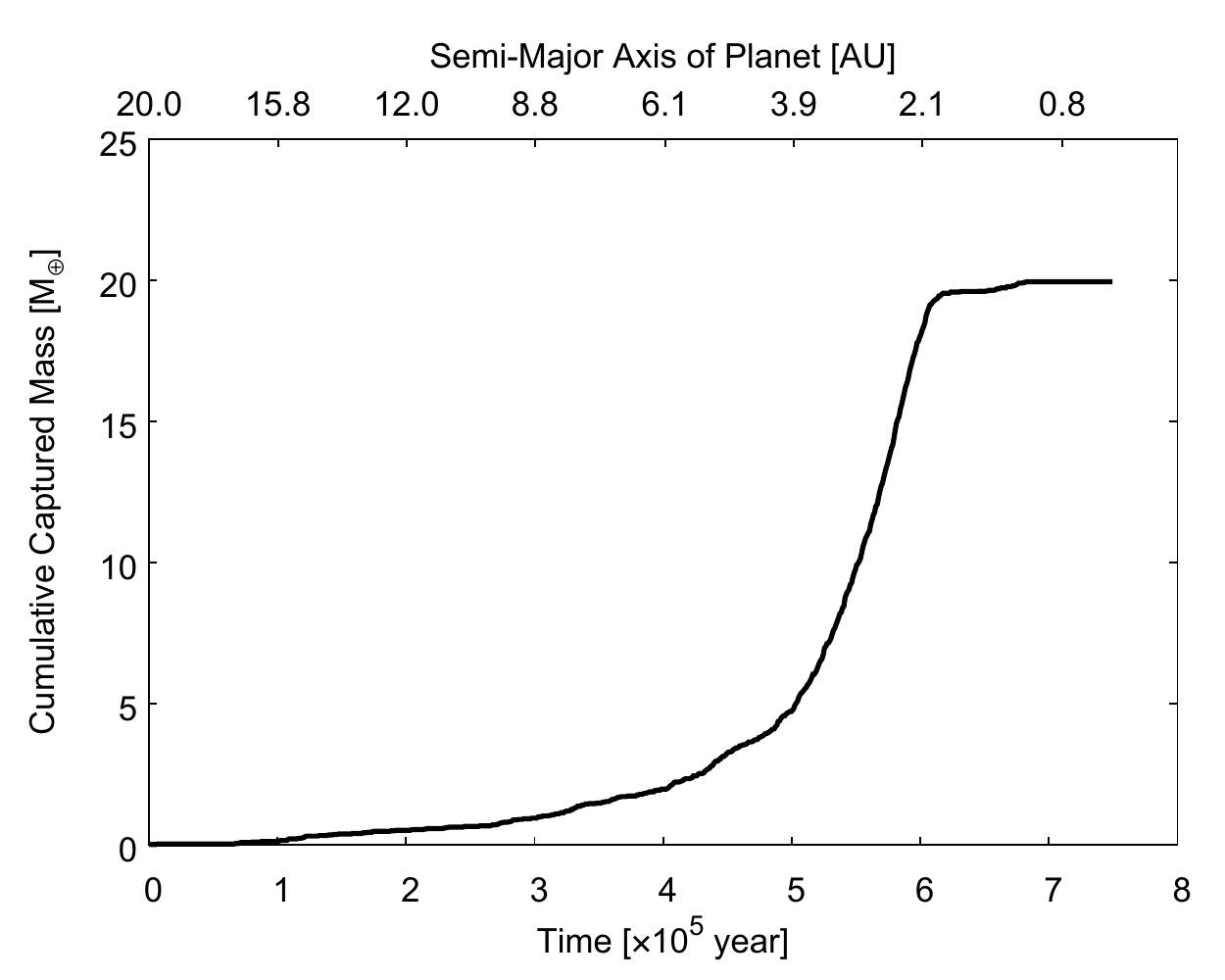}
            \caption{
                The temporal changes in the cumulative mass of captured planetesimals in the reference case (see Table~\ref{tb:settings} for the setting).
                Top x-axis shows the semi-major axis of the migrating planet.
            }
            \label{fig:Result_Cumulative_Captured_Mass}
        \end{center}
    \end{figure}

    Figure~\ref{fig:Result_Snapshots} shows three snapshots of the orbital evolution of planetesimals for the reference case in the semi-major axis vs. eccentricity plane.
    In each panel, the migrating planet is represented by the orange circle, while planetesimals of $E_{\rm J}>0$ and $E_{\rm J}<0$ are indicated by the red and blue circles, respectively.
    The positions of the three mean motion resonances (MMRs) with the planet are indicated by the green-dotted lines. 
    The orbital evolution of planetesimals is summarised below.

    \begin{itemize}
        \item Panel~(a): As the planet migrates inward, planetesimals encounter and are trapped in the MMRs; 
        this phenomenon is referred as resonant trapping.
        The planetesimals trapped in the MMRs are transported inward together with the migrating planet and their eccentricities are highly enhanced.
        This phenomena is known as {\it resonant shepherding} \citep[][]{Batygin+2015}.
        \item Panel~(b): In the course of time, the resonantly-trapped planetesimals start to escape from the MMRs.
        This is because the stronger aerodynamic drag in inner regions allows the planetesimals to escape from the MMRs.
        Such breakup of resonant trapping was found by \citet{Malhotra1993} in the context of the formation of Jupiter's core.
        \item Panel~(c): In the further inner region, the disc gas becomes dense enough that the resonantly trapped planetesimals have their eccentricities damped faster than the planetary migration, and therefore they are outside the feeding zone. 
        We refer to this phenomenon as {\it aerodynamic shepherding} \citep[e.g.][for terrestrial planet formation]{Tanaka+1999}. 
    \end{itemize}

    Figure~\ref{fig:Result_Cumulative_Captured_Mass} shows the temporal changes in the cumulative mass of captured planetesimals.
    The semi-major axis of the migrating planet is shown in the top x-axis. 
    We find that most of the accreted planetesimals are captured mainly during the period between $4~\times~10^5$ and $6~\times~10^5$~years, when the planet migrates from 6 to 2~AU.
    As shown in Fig.~\ref{fig:Result_Snapshots}, the resonant shepherding and the aerodynamic shepherding are significant for $t~\lesssim~4~\times~10^5~\yrs$ and $t~\gtrsim~6~\times~10^5~\yrs$, respectively.
    Most of the planetesimals are captured when the both shepherding processes are inefficient.  
    This result suggests that resonant shepherding and aerodynamic shepherding inhibit planetesimal capture by a migrating giant planet.
    The cumulative captured mass at the end of the calculation (hereafter, the total captured mass $\Mcaptot$) is found to be $\sim20~\Mearth$, which is $\sim 20\%$ of available planetesimals mass  ($\sim100~\Mearth$ planetesimals are distributed at the beginning of the simulation).
    A detail analysis of the shepherding processes is given in Appendix~\ref{App_resonance}.

    \subsection{Dependence on the Initial Semi-Major Axis}\label{sec:Result1_3}
        \begin{figure}
            \begin{center}
            \includegraphics[width=80mm]{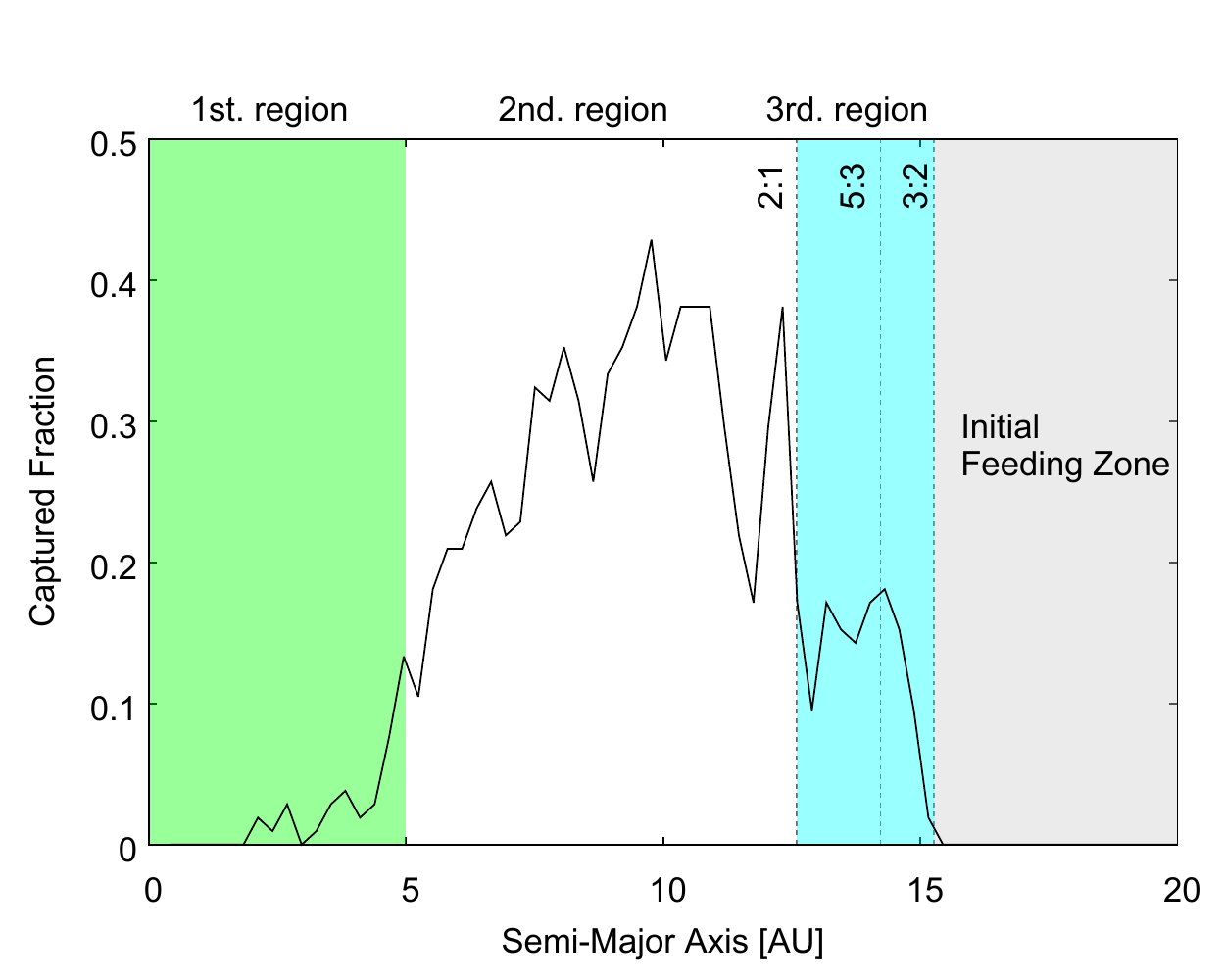}
            \caption{
                The fraction of planetesimals captured by the planet vs.~their source semi-major axis for the reference case.
                The green, white and turquoise  areas correspond to the  first, second, and third regions denoted in Sec~\ref{sec:Result1_3}.
                The difference between these regions is due to the different nature of the dominant shepherding process.
                The gray area shows the initial planetary feeding zone for planetesimals in circular orbits ($e=0$). 
                The dotted lines indicate the positions of the 2:1, 5:3, and 3:2 MMR with the planet at the beginning of the simulation.
                The bin width of the histogram is $1/3~\AU$.
            }
        \label{fig:Result1_Initial_Semi-Major_Axis}
        \end{center}
    \end{figure}
    Figure~\ref{fig:Result1_Initial_Semi-Major_Axis} shows the fraction of captured planetesimals as a function of their source semi-major axis $\axipltsint$.
    From a physical point of view, the histogram can be divided into three regions  including (1) $\axipltsint<5~\AU$, (2) $5~\AU~<\axipltsint<~12.6~\AU$, which corresponds to the initial 2:1 MMR, (3) $12.6 \AU < \axipltsint$.
    In the first region, the fraction is quite small and decreases with decreasing $\axipltsint$.
    Those from the second region are trapped in the 2:1 MMR and, then, about 30-40~\% of them are captured by the planet.
    The planetesimals from the third region are trapped in the 5:3 or 3:2 MMRs and, then, about 10-20~\% of them are captured by the migrating planet.

    In the first region, the dominant shepherding process is the aerodynamic one.
    Since it works more effectively in inner regions, the number of planetesimals that enter the planetary feeding zone decreases with the decreasing initial semi-major axis.

    The difference in capture probability between the second and third regions arises from the difference of the MMRs by which planetesimals are initially trapped.
    As a result, the dynamical configuration of planetesimals inside the feeding zone is changed.
    The main difference is the planetesimal eccentricity; the eccentricity of planetesimals from the third region is higher than that of planetesimals from the second region (see Appendix~\ref{App_resonance}). 
    Since the capture probability decreases with increasing eccentricity \citep{Ida+1989}, the capture probability for the third region is smaller than that for the second one. 

\section{Results of Parameter Study}
\label{sec:Result2}
    \begin{figure*}
        \begin{center}
        \includegraphics[width=160mm]{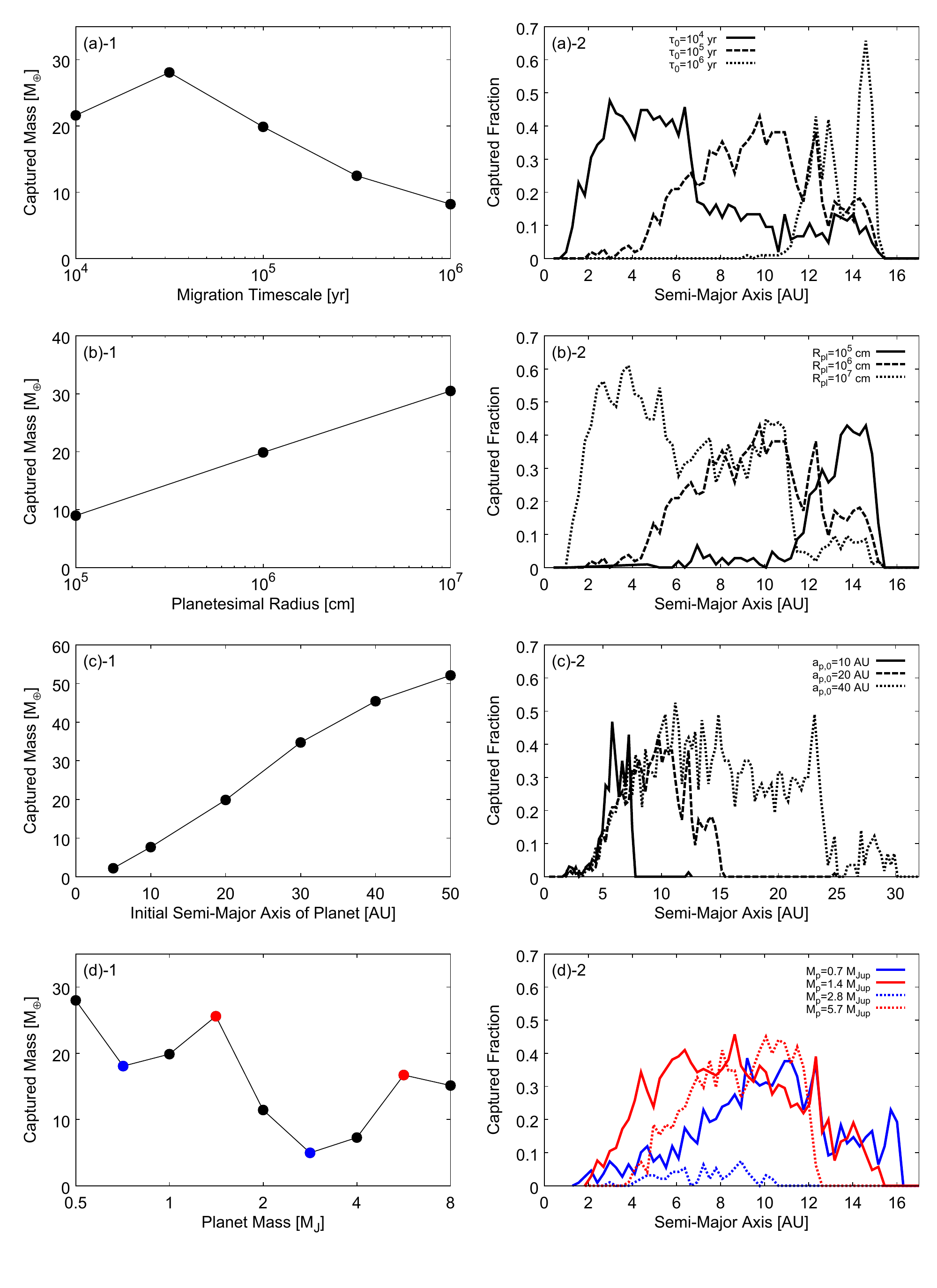}
        \caption{
            Results of the parameter study. \textit{Left column}--the total mass of captured planetesimals as a function of
            (a)-1: the migration timescale of the planet, $\tautidezero$,
            (b)-1: the radius of planetesimals, $\Rplts$,
            (c)-1: the initial semi-major axis of the planet $\axiplanetint$, and
            (d)-1: the mass of the planet $\Mplanet$.
            \textit{Right column}--the fraction of planetesimals captured by the planet as a function of the source semi-major axis (the same as Fig.~\ref{fig:Result1_Initial_Semi-Major_Axis}) for the different cases we consider as indicated in the legend.
            The cases shown with red (blue) plots in (d)-1 are shown with red (blue) lines in (d)-2.
        }
        \label{fig:Result2_Parameter_Study}
        \end{center}
    \end{figure*}

    As discussed above, the two shepherding processes control the efficiency of planetesimal capture by a migrating planet.
    In this section, we perform a parameter study where we change the planet's migration timescale $\tau_{\rm tid,0}$, the radius of planetesimals $R_{\rm pl}$, the migration length (or the initial position of the planet $a_{\rm p,int}$) and planet's mass $M_{\rm p}$, in order to investigate the effects of these parameters on the shepherding and the capture processes of planetesimals.

    Panels in the left column of Fig.~\ref{fig:Result2_Parameter_Study} present the dependence of the total mass of planetesimals captured by the planet, $\Mcaptot$, on ((a)-1) $\tautidezero$, ((b)-1) $\Rplts$, ((c)-1) $a_{\rm p,int}$ and ((d)-1) $\Mplanet$. 
    The right column shows the fraction of planetesimals captured by the planet as a function of their source semi-major axis $\axipltsint$ for the different cases we consider. 

    In panel~(a)-1, we present the values of $\Mcaptot$ calculated for $\log \, (\tautidezero/{\rm yr}) =$ 4.0, 4.5, 5.0, 5.5, and 6.0.
    It is found that $\Mcaptot$ increases with $\tautidezero$ for $10^{4.0} \leq \tautidezero \leq 10^{4.5}$~years, while decreasing with $\tautidezero$ for $\tautidezero \geq 10^{4.5}$~years.
    Two such different trends arise because change in $\tautidezero$ affects both types of shepherding in different ways.
    Aerodynamic shepherding becomes more effective for larger $\tautidezero$, because slower planetary migration makes planetesimals subject to aerodynamic gas drag for a longer time.
    Consequently, as found in panel~(a)-2, the fraction of captured planetesimals from inner regions decreases with increasing $\tautidezero$.
    On the other hand, resonant shepherding becomes less effective for larger $\tautidezero$, because the excitation of planetesimal eccentricity in MMRs weakens with increasing $\tautidezero$.
    As a result of the combination of both shepherdings, $\Mcaptot$ peaks at $\tautidezero=10^{4.5}~\yr$ (see Appendix~\ref{App_resonance} for more details).

    In panel~(b)-1, $\Mcaptot$ is found to increase with $\Rplts$: 
    This is simply because larger $\Rplts$ brings about weaker aerodynamic gas drag and thereby less effective aerodynamic shepherding. 
    Since several studies suggested that planetesimals are likely to born big as $\sim~100~{\rm km}$, an enrichment of several $10 M_{\oplus}$ is expected.
    In panel~(c)-1, $\Mcaptot$ is found to increase with the migration length almost linearly.
    Both shepherding processes are almost independent of the migration length, so $\Mcaptot$ is determined by the number of planetesimals that interact with the planet during its migration.

    Finally, in panel~(d)-1, $\Mcaptot$ is found to change with $\Mplanet$ in a non-monotonic manner and take local maxima at $M_{\rm p}=1.4~M_{\rm J}$ and $5.7~M_{\rm J}$ and local minima at $M_{\rm p}=0.7~M_{\rm J}$ and $2.8~M_{\rm J}$. 
    In panel~(d)-2, the capture fractions are shown for those local maxima and minima cases.
    The reason for such a non-monotonic change in $\Mcaptot$ with $M_{\rm p}$ is linked to the position of MMRs relative to the feeding zone boundary.
    The change in the planet's mass shifts the cross points between the feeding zone boundary and MMRs, because the feeding zone expands with $\Mplanet$ (Eq.~\ref{eq:Ejacobi}).
    Planetesimals enter the feeding zone around these cross points (see Fig.~\ref{fig:Result1_Resonance_Shepherding1} and Fig.~\ref{fig:Result1_Resonance_Shepherding2}): hereafter, these points are referred to as channels.
    The shift of channels is monotonic with $\Mplanet$.
    But the main channel, which supplies most of the planetesimals into the feeding zone, also changes with $\Mplanet$, which results in the non-monotonic change in $\Mcaptot$ with $\Mplanet$.
    The details of this mechanism are presented in Appendix~\ref{App_FlowChannle}.

\section{Discussion}
\label{sec:Discussion}
    \subsection{Implications for Planet Formation}\label{sec:Dis_Compare}
    The total heavy-element mass in a gas giant planet is a combination of the core mass and the heavy elements within the envelope. 
    The core's mass can be thought to be of the order of the critical core mass in the case of a proto-giant planet that forms in outer regions of a relatively massive disc as considered in this study. This is essentially the mass that is accreted during the first phase (phase-1) in the formation of a giant planet. 
    The maximum heavy-element mass that can be accreted by a giant planet forming {\it in-situ} by planetesimal accretion, assuming no planetesimals are accreted during runaway gas accretion, can be given by the isolation mass, $M_\mathrm{iso}$ \citep[][]{Lissauer1987,Kokubo+1998}.
    From the empirical relation for $M_\mathrm{iso}$ obtained by \citet{Kokubo+1998}, $M_\mathrm{iso}$ is 70-80~$\Mearth$ at $20~\AU$ for a solid surface density that is six times higher than the MMSN value.
    However, one should consider that in fact, the critical core can be significantly smaller than the one inferred when assuming dust-rich protoplanetary envelopes \citep{Ikoma+2000,Ikoma+2006b}, due to grain growth and settling which leads to reduction in opacity and shorter formation timescales \citep[][]{Ormel2013,Mordasini2014}.
    In the latter case, the critical core mass is expected to be of the order of $\sim10\Mearth$ or smaller. 
    Even in the case of rapid pebble accretion, pebbles will vaporise in the planetary envelope and therefore significantly reduce the core mass \citep{Hori+2011,Venturini+2015,Brouwers+2018}.
    As a result, the core mass is unlikely to be much higher than 10~$\Mearth$.

    After the first phase, which is dominated by heavy-element accretion, the gas accretion rate becomes much higher than the heavy-element    accretion rate. 
    The dense gaseous envelope prevents planetesimals from reaching the core, and the planetesimals are expected to remain in the envelope.
    In fact, the division between the "core" and "envelope" is not always well-defined \citep{HS2017,Lozovsky2017,Valletta2019}.
    After the onset of the runaway gas accretion, the rapidly expanding feeding zone allows further accretion of planetesimals that are distributed around the planet.
    During that phase, up to $30\%$ of the planetesimals inside the planet's feeding zone can be accreted by the planet \citep{Shibata+2019}.
    As we show here, a higher mass of heavy elements can be captured when planetary migration is considered.

    Using the same parameters as the ones adopted in our reference model, we find that the planet captures $\sim18~\Mearth$ of planetesimals (equivalent to $30~\%$ of all planetesimals inside the Jupiter-mass planet's feeding zone at 20~AU) during the subsequent rapid gas accretion phase.
    Then, during inward migration, the planet captures $20~\Mearth$ of planetesimals (see Sec.~\ref{sec:Result1}).
    Therefore, assuming that a warm-Jupiter that starts its formation at $20~\AU$, it captures $\sim40-50\Mearth$ heavy elements by the end of its migration.

\begin{figure}
      \begin{center}
        \includegraphics[width=80mm]{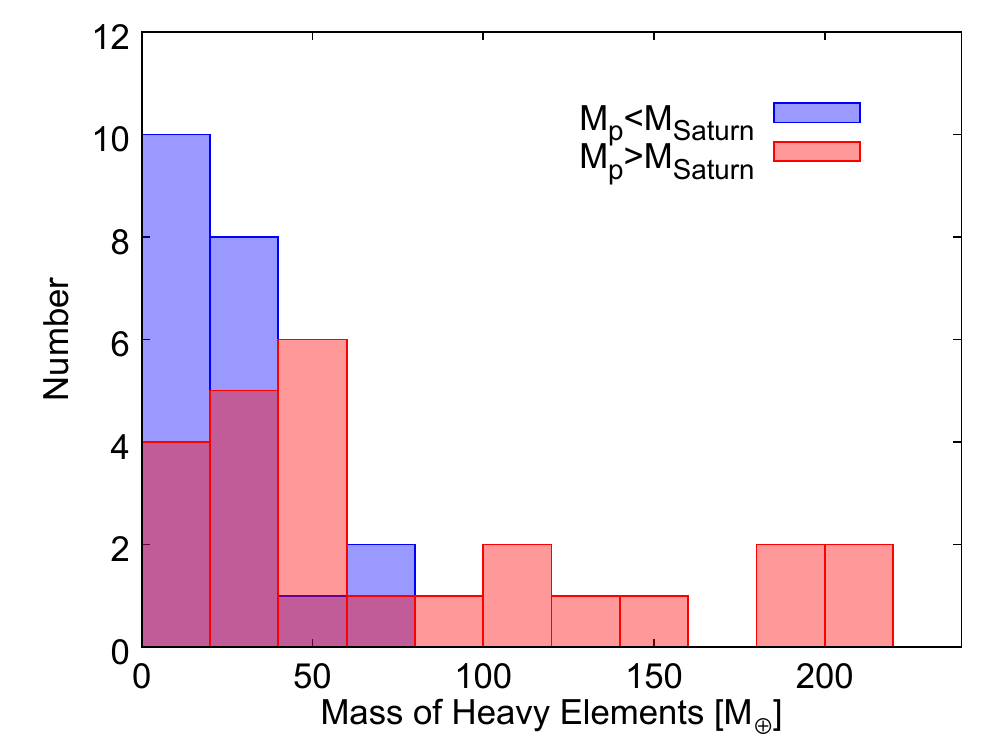}
        \caption{
            The estimated heavy-element mass in warm-Jupiters.
            The histogram is constructed using the results obtained by \citep{Thorngren+2016}.
            The red and blue bars show the planets heavier than Saturn and lighter than Saturn, respectively.
        }
        \label{fig:Thorngren}
        \end{center}
    \end{figure}

    Figure~\ref{fig:Thorngren} shows a histogram of the inferred heavy-element mass in warm-Jupiters, which is made from the data presented in \citet{Thorngren+2016}.
    We divide the data into super-Saturns assuming that these planets have experienced rapid gas accretion, while the sub-Saturns have not.  
    The median of heavy-element mass inferred for these warm-Jupiters (red bars) is $\sim50\Mearth$. 
    This value is consistent with our results. 
    Therefore, our study suggests that the metal-rich warm-Jupiters must have migrated inward from afar to their current location.
    {\it Our study provides an independent way to constrain the formation history of giant exoplanets at small orbits, simply by estimating  their bulk compositions. }
    Our conclusion is in agreement with studies that imply a migration history for these planets based on their orbital parameters \citep[e.g.,][]{Ida+2004a}.
    At the same time, our study emphasises the challenge for planet formation models to explain the highly enriched warm-Jupiters, which contain more than $100~\Mearth$ of heavy elements.

    Our model can explain enrichment of up to $\sim100~\Mearth$ of heavy elements in warm-Jupiters. 
    This is because the amount of heavy elements can be increased when faster planetary migration, larger planetesimals, longer migration lengths (see Fig.~\ref{fig:Result2_Parameter_Study}) or heavier circumstellar discs are considered.
    There are circumstellar discs up to 10-20 times more massive than the MMSN according to observations of young stellar clusters \citep[e.g.,][]{Beckwith+1996,Dullemond+2018} and some of the central star in \citet{Thorngren+2016}'s data have 1-2 times higher metallicity relative to our Sun.
    Nevertheless, even under these idealised assumptions, the super-enriched giant planets cannot be explained. 
    Therefore, we suggest that any further enrichment is expected to be a result of other enrichment mechanisms such as giant impacts and planet merging \citep[][]{Ikoma+2006b,Liu2015} or additional processes neglected in our model, as discussed below. 

    Our study suggests that the relation between the heavy-element mass $M_{\rm p,Z}$ and planet mass $\Mplanet$ found by \citet{Thorngren+2016} is rather complex.
    We show that there are no simple relation between $\Mcaptot$ and $\Mplanet$ (Fig.~\ref{fig:Result2_Parameter_Study}(d)-1).
    On the other hand, $\Mcaptot$ increases almost linearly with the initial position of the planet $a_{\rm p,int}$ (Fig.~\ref{fig:Result2_Parameter_Study}(c)-1), suggesting that the formation of more enriched giant exoplanets starts at outer region of the disc.
    Therefore, our results predict that the relation between $M_{\rm p,Z}$ and $\Mplanet$ depends on the relation between the mass of gas giant planets and the position where runway gas accretion starts, which is consistent with the results of \citet{Tanigawa+2007} and \citet{Tanigawa+2016}.

    Recently, \cite{Humphries+2018} considered giant planet formation via gravitational instability and accretion of pebbles during the rapid type 1 migration in order to reproduce the trends found by \citet{Thorngren+2016}.
    Their results imply that the planetary  metallicity is increased by pebble accretion because rapid migration hinders gap opening in the pebble disc.
    Our model cannot be directly compared with this work because we focus on type 2 migration.
    Nevertheless, our results suggest that the gas giants migrating with type 1 regime can also capture many heavy elements even in a disc in which the solid material is in the form of planetesimals. 
    In fact, heavy-element enrichment of clumps formed by gravitational instability via planetesimal capture was also found to be efficient and to lead to enrichments of several tens $M_{\oplus}$, depending on the planetary formation location and disc properties \citep{Helled09}. 
    We suggest that future studies should compare the different predictions for the heavy-element enrichment of gaseous planets considering both planetesimal and pebble accretion, type 1 and type 2 migration, and formation via both core accretion and disk instability. 

    Finally, it should be noted that in terms of planetary bulk metallicity, migration is much more significant for the lower planetary masses; for instance, for a 0.5 $M_J$ with an initial core mass of 10 $M_{\oplus}$, the total heavy-element mass can increase by a factor of four, resulting in a planetary metallicity of $\sim 30\%$.
    Also, it is important to keep in mind that the inferred heavy-element masses of the giant exoplanets can vary depending on the model assumptions, in particular, the used equation of state for hydrogen and helium, the gas and dust opacity, the assumed compositions and thermal profile, etc. 
    It is therefore desirable to investigate the sensitivity of the inferred planetary metallicity of warm/hot Jupiters in a more rigorous way, and we hope to address this in future research.

    \subsection{Resonant Braking in Planet-Planetesimal Resonance}\label{sec:Dis_Res_Brk}
    Planetesimals escape from MMRs during planetary migration  because strong aerodynamic gas drag amplifies the libration of resonance angle and the libration of semi-major axis exceeds the resonant width of MMRs (see Sec.~\ref{App_resonance} for details). 
    This mechanism was named {\it overstable libration} in the context of planet-planet resonant breaking \citep{Goldreich+2014,Hands+2018}. 
    The main difference between their planet-planet resonant breaking and our planet-planetesimal resonant breaking is the source of damping. 
    The planet's orbit is damped by the gravitational interaction with the circumstellar disc, while the planetesimal's orbit is damped by the aerodynamic interaction with circumstellar disc.
    We find that this "overstable libration" occurs even in planet-planetesimal resonance because the planetesimal's eccentricity reaches a very high value ($\sim~0.1$) leading to the damping timescale of the eccentricity to be shorter than that of the planetary migration, which is a requirement of this resonant breaking.
    Our results show that overstable libration plays an important role for planetesimals accretion.
    A detailed analysis of planet-planetesimal resonant breaking is desirable, but is beyond the scope of this paper, and we hope to address it in future research.

    The results of our parameter study regarding the planetary mass (see Fig.~\ref{fig:Result2_Parameter_Study}) might be important for understanding the  planet-planetesimal resonance and its resonant breaking. 
    This unique relation between heavy-element mass and planet mass can be observable, assuming there are no other effects that lead to the non-monotonic relation between them.
    The planetary mass-enrichment dependency derived in this study is not found in the sample of  \citet{Thorngren+2016}, however, this could be a result of limited data. 
    As a result, we suggest that further observations could be used to better understand the role of planet-planetesimal resonance in the capture of planetesimals by a migrating planet.

    \subsection{Effects neglected in this work}\label{sec:Dis_Mod_Eff}
    In this study, we have considered only one migrating planet.
    Also we have treated the planetesimals as test particles and ignored their mutual collisions and gravitational forces on the migrating planet. 
    The effects of the accreted solids on the planetary evolution (via changes in composition and opacity) are also neglected.
    In addition, we have used Eq.~(\ref{eq:Rplanet_cap}) assuming a constant mean density $\rhoplanet~=0.125~\g~\cm^{-3}$ for the planet's radius, although the planetary capture radius can differ from this value \citep[e.g.][]{Podolak+1988,Inaba+2003}.
    Below, we discuss the potential effects of these processes on our results.

    When a gas giant planet starts to migrate, many other planets could exist interior to the orbit of the gas giant in the same system.
    If the relative distance between the gas giant planet and other planets decreases due to their different migration timescales (converging process), those planets ahead of the gas giant planet could scatter planetesimals into the giant planet's feeding zone, reducing the efficiency of aerodynamic shepherding.
    In addition, the gravitational perturbation from those planets on planetesimals trapped in MMRs would amplify the libration of their orbits, thereby breaking the resonant shepherding process.
    Thus, in the converging processes, the existence of other planets is expected to enhance the efficiency of capture of planetesimals by the migrating giant planet. 
    On the other hand, if the relative distance increases (diverging process), planetesimals would be swept by those planets, reducing the heavy-element mass accretion efficiency. 
    These effects should be investigated in future research.

    During the migration process, the surface density of planetesimals around MMRs increases. 
    The timescale of collision between the planetesimals becomes shorter with increasing surface density.
    The collision between planetesimals results in two opposite effects on the planetesimal capture efficiency.
    First, collisions perturb the planetesimals' orbits and some of the collisions can eject planetesimals from resonant trapping.
    \citet{Malhotra1993} investigated this effect and found that the velocity and direction of collision determines whether the resonant trapping is broken.
    This effect increases the efficiency of planetesimal capture by a migrating giant planet.
    Second, collisions between highly eccentric planetesimals lead to shattering, which produces many small fragments.
    This effect reduces the planetesimal capture efficiency, because small size planetesimals are easily shepherded by a migrating planet (see \ref{fig:Result2_Parameter_Study}).
    Thus, collisions between planetesimals have such effects of increasing and decreasing the heavy-element enrichment of a migrating giant planet; the net effect is our future investigation.

    Due to the shepherding processes, many planetesimals are pushed inward by a migrating planet.
    At the same time, these planetesimals push the migrating planet in the opposite direction; this effect is neglected in our model.
    The maximum amount of planetesimals being shepherded by the migrating planet is $\sim50~\Mearth$ in the reference model, which corresponds to  $\sim20\%$ of the mass of the migrating planet.
    Due to the exchange of torques, the migration timescale would be prolonged by $\sim20\%$ \citep{Ida+2004a}.
    Therefore, such effect slows down the migration speed of the planet.

    After gas accretion terminates, giant planets contracts gradually.
    The high accretion rate of heavy-elements, however, affects the planetary evolution due to the change in the planetary composition, thermal energy, and opacity.
    While the accreted heavy-elements can lead to an expansion of the radius due to the additional energy and increased (gas+dust) opacity, it can also decrease the radius due to the addition of heavy elements.
    There is also a possibility that the high heavy-element accretion rate would results in an increase in luminosity which in return may decelerate the planetary contraction.
    The expansion of the planet radius results in the enhancement of capture radius, and enhancement of capture rate of planetesimals.
    The feedback effect occurs between the capture of planetesimals and the planet radius.
    It is therefore desirable to perform a calculation of the planetary evolution taking into account the migration and heavy-element accretion self-consistently, and we hope to address this in future research.

\section{Summary}
\label{sec:Summary}
    A significant fraction of the warm-Jupiters with measured masses and radii are thought to be metal-rich relative to their host stars. 
    Moderate heavy-element enrichment can be explained by planetesimal capture in the late-formation stages of gas giant planets. 
    In this study, we have performed numerical simulations of the dynamics of planetesimals around a migrating giant planet, taking into account the effects of mean motion resonances, and investigated the fundamental physics for planetesimal capture during planetary migration. 
    We have then investigated the total amount of planetesimals that the giant planet finally captures for several model parameters including the migration timescale, the size of planetesimals, the migration length, and the planetary mass.
    Our main findings can be summarised as follows:

    \begin{itemize}
        \item There are two shepherding processes, resonant shepherding and aerodynamic shepherding, that act as barriers for planetesimal capture by a migrating gas giant planet.
        \item Aerodynamic gas drag has two key effects that lead to breaking the resonant shepherding and, on the other hand, causing aerodynamic shepherding.
        \item Planetesimal capture occurs when both shepherding processes are ineffective.
        \item A migrating Jupiter-mass planet can capture planetesimals with total mass of a few tens of Earth masses, provided it starts migration at a few tens of AU in a relatively massive planetesimal disc.
        \item The planetesimal capture efficiency peaks at the moderate migration timescale $\tau_{\rm tid,0}=10^{4.5}\yr$.
        \item A migrating planet captures larger planetesimals more efficiently.
        \item The captured heavy-element mass increases with the migration length almost linearly.
        \item The planetesimal capture efficiency depends on the position of MMRs relative to the feeding zone boundary.
    \end{itemize}

    We conclude that planetary migration increases planetary metallicity, and suggest that the heavy elements in warm/hot-Jupiters are mainly brought by planetesimal accretion during planetary migration, and that the relation between the heavy-element mass and planetary mass could be explained by different migration distances. 

    Clearly, our work has not included all the governing physical processes, and there is much more work to be done in the future, including the effects of the existence of other planets, collision among planetesimals, pebble accretion, and changes in internal structure of the gas giant planet.
    Therefore our study should be considered as a first step in a detailed investigation of the enrichment mechanisms of gas giant planets.
    This will provide a better understanding of the gas giants in the solar systems as well as of giant exoplanets.

\newpage
\appendix
\section{Drag Coefficient}\label{App_Cd}
    In general, the non-dimensional drag coefficient $\cd$ is a function of the Reynolds number $\Reyn$ (= $2 \rhogas \Rplts \vplgs  / \mu$; $\vplts$ being velocity of planetesimals and $\mu$ being dynamic viscosity) and the Mach number $\Mach$ (= $\vplts / \csound$; $\csound$ being the sound speed).
    The dynamic viscosity is given by $\mu = (1/3) \rhogas \csound \lmfp$, where $\lmfp$ is the mean free path of gas molecules for which we adopt the collision cross section of hydrogen molecules (= $2 \times 10^{-15} {\cm}^2$).
    In this study, we use an approximated formulae for $\cd$ written as \citep[e.g.][]{Tanigawa+2014}
    \begin{align}\label{eq:Drag_Coefficient_Tanigawa}
	    \cd \simeq \left[ \left( \frac{24}{\Reyn} + \frac{40}{10+\Reyn} \right)^{-1} + \frac{3 \Mach}{8} \right]^{-1} + \frac{ (2-\Ocrr) \Mach}{1+\Mach} + \Ocrr,
    \end{align}
    where $\Ocrr$ is a correction factor, the value of which is 0.4 for $\Reyn < 2 \times 10^5$ and 0.2 for $\Reyn > 2 \times 10^5$. 

\section{Gap Structure in a Circumstellar Disc}\label{App_disk}
    \cite{Kanagawa+2017} derived an empirical formulae for the gas structure based on the results of hydrodynamic simulations.
    They found that there is a relation between the gap width $\Delta_{\rm gap}$ and non-dimensional parameter $K^{\prime}$ as
    \begin{align}
        \frac{\Delta_{\rm gap} \left(\Sigma_{\rm th}\right)}{R_{\rm p}} = \left( 0.5 \frac{\Sigma_{\rm th}}{\Sigma_{\rm un}} +0.16 \right) {K^{\prime}}^{1/4},
    \end{align}
    where $\Sigma_{\rm th}$ and $\Sigma_{\rm un}$ are the surface density of disc gas inside the gap and outside the gap, respectively, and $K^{\prime}$ is a non-dimensional parameter defined as
    \begin{align}\label{eq:gap_Kp}
        K^{\prime} = \left( \frac{M_{\rm p}}{M_{\rm s}} \right)^2 \left( \frac{h_{\rm s}}{a_{\rm p}} \right)^{-3} {\alpha_{\rm \nu}}^{-1}.
    \end{align}
    They also found the surface density at the gap bottom as
    \begin{align}
        \frac{\Sigma_{\rm min}}{\Sigma_{\rm un}} = \frac{1}{1+0.04K},
    \end{align}
    where $K$ is the non-dimensional parameter defined as
    \begin{align}\label{eq:gap_K}
        K = \left( \frac{M_{\rm p}}{M_{\rm s}} \right)^2 \left( \frac{h_{\rm s}}{a_{\rm p}} \right)^{-5} {\alpha_{\rm \nu}}^{-1}.
    \end{align}
    Using above results, they constructed an empirical function representing the radial distribution of the gap structure $f_{\rm gap}$ as 
    \begin{align}\label{eq:surface_density_gap}
        f_{\rm gap} = \left\{
        \begin{array}{ll}
            \frac{1}{1+ 0.04 K}                                             & \text{for $| r - \axiplanet | < \Rone$}, \\
            4.0 {K^{\prime}}^{-1/4} \frac{|r-\axiplanet|}{\axiplanet} -0.32 & \text{for $\Rone < | r - \axiplanet | < \Rtwo$}, \\
            1                                                               & \text{for $\Rtwo < | r - \axiplanet |$},
        \end{array}
        \right.
    \end{align}
    where
    \begin{align}
	    \Rone   &= \left( \frac{1}{1+ 0.04 K} + 0.08 \right) {K^{\prime}}^{1/4} \axiplanet, \label{eq:gap_r1}\\
	    \Rtwo   &= 0.33 {K^{\prime}}^{1/4} \axiplanet. \label{eq:gap_r2}
    \end{align} 

\section{A Detailed Analysis of Shepherding Processes}\label{App_resonance}
    \begin{figure*}
        \begin{center}
        \includegraphics[width=170mm]{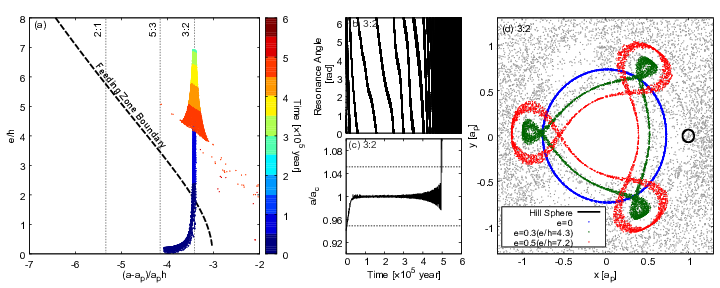}
        \caption{
            Orbital evolution of a single planetesimal initially located at the semi-major axis $a_{\rm p,0}$ = 14.3~AU.
            Panel~(a) shows the evolutionary path of the planetesimal in the plane whose vertical and horizontal axes are the eccentricity, $e$, and the difference in semi-major axis between the planetesimal, $a$ and planet, $a_\mathrm{p}$, respectively; both are normalised by the reduced Hill radius, $h$ (see Eq.~[\ref{eq:reduced_Hill_radius}]). 
            The time sequence is colour-coded.
            The dashed line indicates the boundary of the feeding zone.
            The dotted lines are eye-guide ones that show the resonance centre of the 2:1, 5:3 and 3:2 mean motion resonances (MMRs) with the planet, from left to right.
            Panels~(b) and (c) show the temporal changes in the resonance angle (see Eq.~[\ref{eq:resonance_angle}]) and in the semi-major axis of the planetesimal normalised by that of the resonance centre, $a_c$ (see Eq.~[\ref{eq:resonance_width}]), respectively, for the 3:2 MMR.
            The dotted lines indicate the resonant width of the 3:2 MMR.
            Panel~(d) shows the orbits of the planetesimal on the co-rotating flame with the migrating planet.
            The blue, green and red dots show the path of the planetesimal for $e=0$, $0.2$ and $0.4$, respectively.
            The gray dots show the orbit of the planetesimal after escaping from the resonance. 
            The black circle shows the Hill sphere of the planet.
            }
        \label{fig:Result1_Resonance_Shepherding1}
        \end{center}
    \end{figure*}

    \begin{figure*}
        \begin{center}
        \includegraphics[width=170mm]{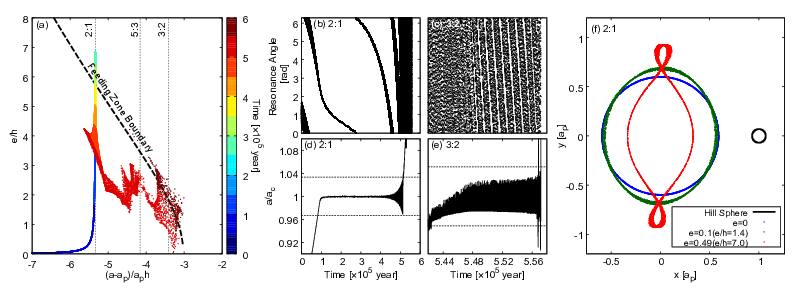}
        \caption{
            Orbital evolution of a single planetesimal initially located at $a_{\rm pl,0}$ = 10.0~AU.
            Panel~(a) is the same as Fig.~\ref{fig:Result1_Resonance_Shepherding1}a, but for $a_{\rm pl,0}$ = 10.0~AU.
            Panels (b)-(c) and (d)-(e) show the temporal changes in the resonance angle (see Eq.~[\ref{eq:resonance_angle}]) and in the planetesimal's semi-major axis relative to the resonance centre, $a_c$ (Eq.~[\ref{eq:resonance_width}]), respectively, in the 2:1 and 3:2 mean motion resonances (MMRs) with the planet. 
            Panel~(f) is also the same as Fig.~\ref{fig:Result1_Resonance_Shepherding1}d, but for the planetesimal with $e$ = 0, 0.1, and 0.5 in the 2:1 MMR. 
            See the caption of Fig.~\ref{fig:Result1_Resonance_Shepherding1} for the details.
            }
        \label{fig:Result1_Resonance_Shepherding2}
        \end{center}
    \end{figure*}

    In Sec.~\ref{sec:Result1}, we have found that two shepherding processes play important roles in the process of planetesimal capture by the migrating planet. 
    In this section, we analyse those shepherding processes and their influence on the efficiency of the planetesimal capture in detail.

\subsection{Resonant Shepherding}\label{sec:Result1_2_1}
    First, the resonant trapping requires the following conditions: 
    \begin{itemize}
        \item[(i)] the eccentricity of the planetesimal on the approach to resonance is smaller than a critical one $e_{\rm crit}$,
        \item[(ii)] the libration width of the planetesimal orbit is smaller than the resonance width.
    \end{itemize}
    The critical eccentricity for the ($p$+$q$):$p$ MMR is given by \citep[e.g.][]{Murray+1999}
    \begin{align}
        e_{\rm crit} = \sqrt{6} \left\{ \frac{3}{\zeta_{p,q} } p^{4/3} (p+q)^{2/3} \frac{M_{\rm s}}{M_{\rm p}} \right\}^{-1/3},
    \end{align}
    where $\zeta_{p,q}$ is the interaction coefficient, the values of which are $-1.19$, $-2.02$, and $3.27$ for the 2:1, 3:2, and 5:3 MMRs.
    In the reference case, $e_{\rm crit}$ = $0.15$, $0.12$ and $0.11$ for the 2:1, 3:2, and 5:3 MMRs, respectively.
    The resonance width is given in the terms of the semi-major axis as \citep[e.g.][]{Murray+1999}
    \begin{align}\label{eq:resonance_width}
        \left|\frac{\Delta a}{a_{\rm c}}\right|_{\rm lib} = 4 \left\{ \frac{1}{3} \frac{M_{\rm p}}{M_{\rm s}} \left(\frac{p}{p+q}\right)^{2/3} \zeta_{p,q} e\right\}^{1/2},
    \end{align}
    where $a_{\rm c}$ is the semi-major axis of the resonance's centre, which are $0.063e^{1/2}$, $0.091e^{1/2}$, and $0.11e^{1/2}$ for the 2:1, 3:2 and 5:3 MMRs, respectively.

    The motion of test particles trapped in the ($p$+$q$):$q$ MMR is known to have an adiabatic invariant given by \citep{Yu+2001}
    \begin{align}\label{eq:Adiabatic_Invariant}
        \sqrt{M_{\rm s} a} \left\{ (p+q)-p\sqrt{1-e^2} \cos i \right\} = \mathrm{constant}, 
    \end{align}
    provided the particles are exerted on by no non-conservative force such as gas drag.
    By taking the time derivative of Eq.~(\ref{eq:Adiabatic_Invariant}), one obtains the excitation rate of the eccentricity as
    \begin{align}\label{eq:Adiabatic_Invariant_deriv}
        \left.\deriv{e}{t}\right|_{\rm ad} &= \frac{\sqrt{1-e^2}}{2ep} \left\{(p+q)-p\sqrt{1-e^2}\right\} \tau_{\rm tid,a}^{-1}
    \end{align}
    for $i = 0$.
    During resonant trapping, the planetesimal's eccentricity is excited by the planet's gravitational scattering and damped by the aerodynamic gas drag. 
    The above equation indicates that the excitation rate decreases with increasing eccentricity, while the damping rate increases with increasing eccentricity \citep{Adachi+1976}.
    As a result, the eccentricity of planetesimals takes an equilibrium value during the resonant trapping.

    Whether a test particle is trapped in a MMR is often checked by use of the resonance angle defined as
    \begin{align} \label{eq:resonance_angle}
        \phi = (p+q) \lambda_{\rm p} -p \lambda -q \varpi,
    \end{align}
    where $\lambda_{\rm p}$ and $\lambda$ are the mean longitude of the planet and the planetesimal, respectively, and $\varpi$ is the longitude of pericenter of the planetesimal.
    While circulating rapidly outside MMRs, the resonance angle librates inside MMRs.

    Figure~\ref{fig:Result1_Resonance_Shepherding1} shows the orbital evolution of a planetesimal initially located at 14.3~AU.
    In panel~(a), the evolutionary path of the planetesimal is shown on the $\tilde{b}-\tilde{e}$ plane, where
    \begin{align}
        \tilde{b} \equiv \frac{a-a_{\rm p}}{a_{\rm p} h}, \,\,\,\,\, \tilde{e} \equiv \frac{e}{h}.
    \end{align}
    On this plane, the boundary of the feeding zone (dashed line) and the locations of the resonance centres (dotted lines) are fixed in our model.
    At the beginning of the simulation, the planetesimal is in between the 5:3 and 3:2 MMRs.
    As the planet migrates inward, the planetesimal is trapped in the 3:2 MMR, gets its eccentricity highly excited, and then enters the feeding zone.
    The eccentricity keeps increasing until it reaches the equilibrium value mentioned above ($\tilde{e}\sim~7.0$).

    In panels~(b) and (c), the resonance angle (Eq.~[\ref{eq:resonance_angle}]) and the the planetesimal's semi-major axis relative to the resonance centre $a_c$ are shown with time.
    In panel (c), the resonance width given by Eq.~(\ref{eq:resonance_width}) is also shown with the dotted lines; its eccentricity is substituted from the simulation's result.
    When $t\lesssim4\times10^5\yr$, the resonance angle circulates relatively slowly from 0 to $2\pi$ (Fig.~\ref{fig:Result1_Resonance_Shepherding1}b) and the libration width is much smaller than the resonance width (Fig.~\ref{fig:Result1_Resonance_Shepherding1}c).
    After the eccentricity reaches the equilibrium value, however, the libration is amplified due to the strong aerodynamic gas drag.
    Once the libration width exceeds the resonance width, the resonance angle starts to circulate rapidly, breaking the resonant trapping.

    Panel~(d) shows the planetesimal's orbits on the co-rotating frame with the migrating planet before (red, green and blue dots) and after (gray dots) escaping from the 3:2 MMR.
    It turns out that during the resonant trapping, the planetesimal never enters the planet's Hill sphere (indicated by a black line) even inside the feeding zone.
    Once escaping from the resonance, however, the planetesimal experiences many close encounters with the planet, ending up being captured by the planet or eliminated from its feeding zone.

    The orbital evolution of a planetesimal initially located somewhat closer to the central star is shown in Fig.~\ref{fig:Result1_Resonance_Shepherding2}, which is the same as Fig.~\ref{fig:Result1_Resonance_Shepherding1} but for $\axipltsint$ = $10~\AU$. 
    As shown in Fig.~\ref{fig:Result1_Resonance_Shepherding2}a, in contrast to the case shown in Fig.~\ref{fig:Result1_Resonance_Shepherding1}a, the planetesimal once enters the feeding zone, but then gets out from the feeding zone through the 2:1 MMR.
    This is because the transport of the resonantly trapped planetesimal to inner high-density region leads to lowering the equilibrium eccentricity to below the eccentricity corresponding to the feeding zone boundary before the resonant trapping is broken.
    After escaping from the 2:1 resonant trapping with $\tilde{e}\sim4$, the planetesimal undergoes other resonant trappings.
    Panels~(b)-(c) and (d)-(e) show the resonance angle and semi-major axis (relative to the resonance centre) of the planetesimal in the 2:1 and 3:2 MMRs, respectively.
    We find that the planetesimal escaping from the 2:1 resonant trapping is trapped in the 3:2 MMR.
    After staying in the 3:2 MMR for a while, the planetesimal ends up entering the feeding zone.
    Because of the same reason as the 3:2 resonant trapping (Fig.~\ref{fig:Result1_Resonance_Shepherding1}d), the planetesimal never enters the Hill sphere during being trapped in the 2:1 MMR, as shown in panel~(f).

\subsection{Aerodynamic Shepherding}\label{sec:Result1_2_2}
    \begin{figure}
        \begin{center}
        \includegraphics[width=80mm]{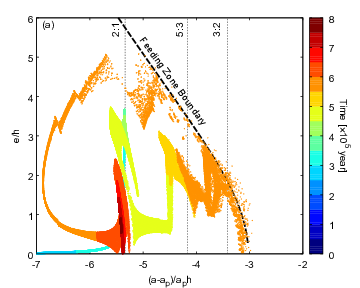}
        \caption{
            Same as Fig.~\ref{fig:Result1_Resonance_Shepherding1}a, but for $\axipltsint=5.0\AU$.
        }
        \label{fig:Result1_Aerodynamic_Shepherding}
        \end{center}
    \end{figure}

    Inward migration of the planet (or a decrease in distance between the planet and planetesimal) leads to increasing the Jacobi energy of the planetesimal orbiting interior to the planet's orbit, whereas the aerodynamic gas drag reduces it.
    As the planet migrates inward, the timescale of migration slowly decreases ($\propto{\axiplanet}^{1/2}$; see Eq.~[\ref{eq:migration_timescale2}]), while that of damping by aerodynamic gas drag decreases even faster because the disc gas density depends on $\axiplts$ more strongly than ${\axiplts}^{1/2}$.
    In the reference model, the latter overwhelms the former at $\axiplanet~\lesssim$~2~AU. 
    Indeed, in the case of Fig.~\ref{fig:Result1_Aerodynamic_Shepherding}, which shows the evolution path of a planetesimal initially located at $5.0\AU$, for example, the aerodynamic gas drag is so strong when the planetesimal reaches the feeding zone boundary (the planet orbiting at 3~AU) that the aerodynamic shepherding inhibits the planetesimal from entering the feeding zone.

\subsection{Effects of Resonant and Aerodynamic Shepherding Processes on Capture of Planetesimals}\label{sec:Result1_2_3}
    \begin{figure}
        \begin{center}
        \includegraphics[width=80mm]{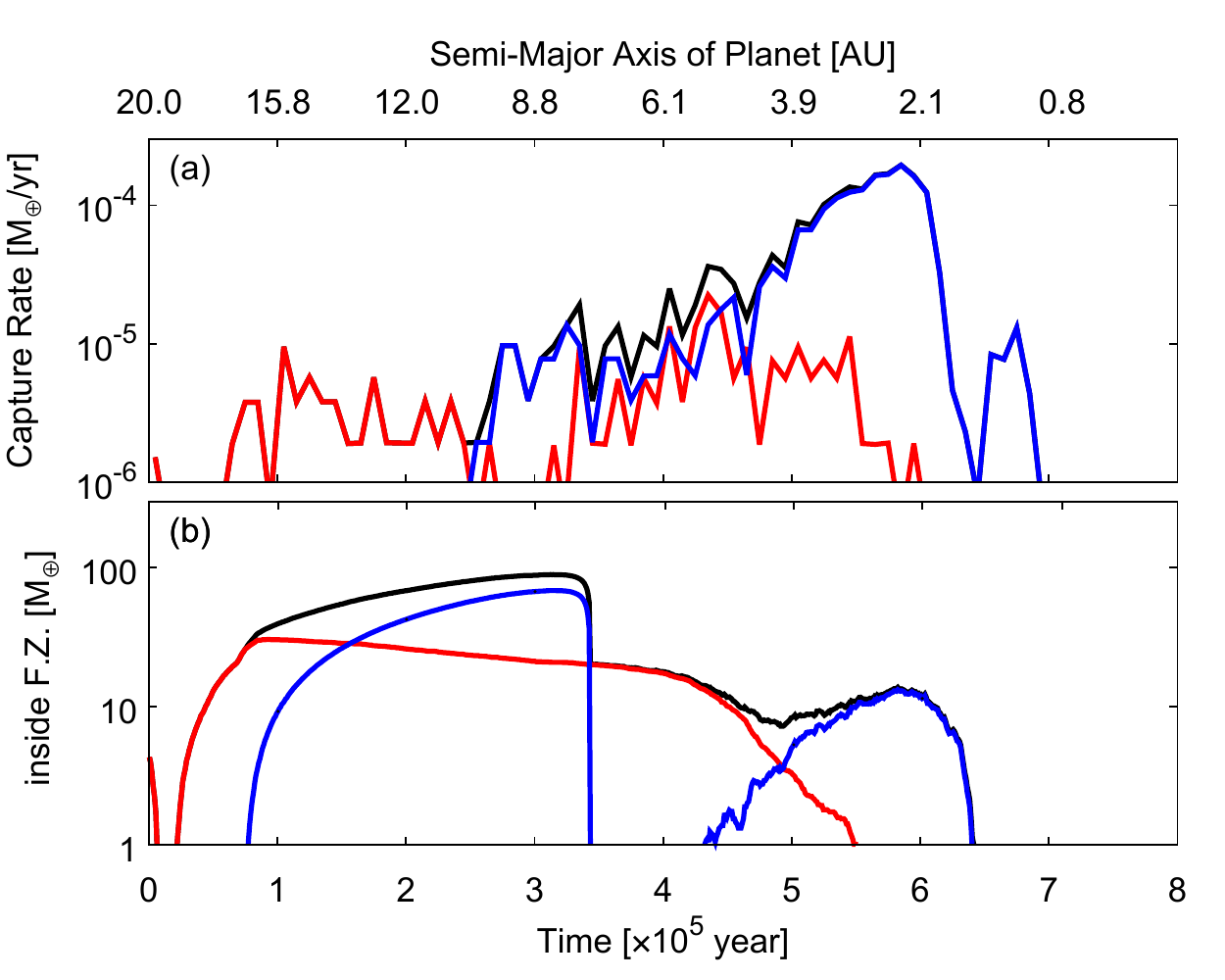}
        \caption{
            The temporal changes in (a) the capture rate of planetesimals and (b) the total mass of planetesimals inside the feeding zone in the reference case (see Table~\ref{tb:settings} for the setting).
            The red and blue lines show the contributions of planetesimals initially located on the side far from and close to the central star relative to the initial position of the 2:1 mean motion resonance (12.6~AU), respectively; the black one shows all the contributions.
            Top x-axis shows the semi-major axis of the migrating planet. 
        }
        \label{fig:dMcap_dt_reference}
        \end{center}
    \end{figure}

    Using the analysis of the resonant shepherding and aerodynamic shepherding presented above, we next  reanalyse the results of Sec.~\ref{sec:Result1}.

    Figure~\ref{fig:dMcap_dt_reference} shows the temporal changes in (a) the planetesimal capture rate and (b) the total mass of planetesimals inside the planet's feeding zone. 
    The red and blue lines show the contributions of planetesimals initially located on the side far from and close to the central star relative to the initial position of the 2:1 MMR with the planet (12.6~AU), respectively; the black line shows the sum of both contributions. 
    As can be seen from Fig.~\ref{fig:Result_Cumulative_Captured_Mass}a and b, it is found that the planetesimal capture rate correlates with the total mass of planetesimals inside the feeding zone for $t~\gtrsim~4~\times~10^5~\yrs$, but this dependence is not observed for $t~<~4~\times~10^5~\yrs$.
    We also find that the total mass of planetesimals inside the feeding zone rapidly increases around $t~\sim~1~\times~10^5~\yrs$ and suddenly decreases at $t~\sim~3~\times~10^5~\yrs$.
    These features are related with the resonant shepherding by the 2:1 and 3:2 MMRs. 

    The rapid increase of total planetesimal mass inside the feeding zone is brought by the resonant shepherding as shown in Fig.~\ref{fig:Result1_Resonance_Shepherding1}a and Fig.~\ref{fig:Result1_Resonance_Shepherding2}a.
    As the planet migrates inward, the equilibrium eccentricity at the 2:1 MMR, which was initially inside the feeding zone in the $\tilde{b}$-$\tilde{e}$ plane, becomes smaller than the eccentricity corresponding to the feeding zone boundary. 
    Then, the total mass of planetesimals inside the feeding zone suddenly decreases at $t~\sim~3~\times~10^5~\yrs$.
    When $t~<~4~\times~10^5~\yrs$, planetesimals inside the feeding zone are trapped in the 2:1 or 3:2 MMRs and on "out-of-phase" librating orbits that never approach the planet's Hill sphere (see Fig.~\ref{fig:Result1_Resonance_Shepherding1}d and Fig.~\ref{fig:Result1_Resonance_Shepherding2}f), thus the planetesimal capture rate is low despite the large number of planetesimals within the feeding zone.

    As explained in Sec.~\ref{sec:Result1_2_1}, the planetesimals start to escape from the resonant traps at $t\sim4\times10^5$~years; some of the planetesimals that were in the 3:2 MMR from early on are captured by the planet (the red line in Fig.~\ref{fig:Result_Cumulative_Captured_Mass}) and some of the planetesimals escaping from the 2:1 MMR enter the feeding zone through the 3:2 MMR (the blue line), which further increases the total mass of planetesimals inside the planet's feeding zone until $t\sim6\times10^5$~years.
    Finally, the aerodynamic shepherding hinders planetesimals from entering the planet's feeding zone.
    By the time the planet reaches $\sim1~\AU$, almost all the planetesimals are eliminated from the feeding zone because of the aerodynamic shepherding.
    Our analysis demonstrates that the resonant shepherding delays the main accretion phase until the planet reaches $\sim 6~\AU$, and the aerodynamic shepherding stops planetesimal accretion when the planet reaches $\sim~1~\AU$.

\section{Chanel of Planetesimal Flow into Planetary Feeding Zone}\label{App_FlowChannle}
    \begin{figure}
    \begin{center}
        \includegraphics[width=80mm]{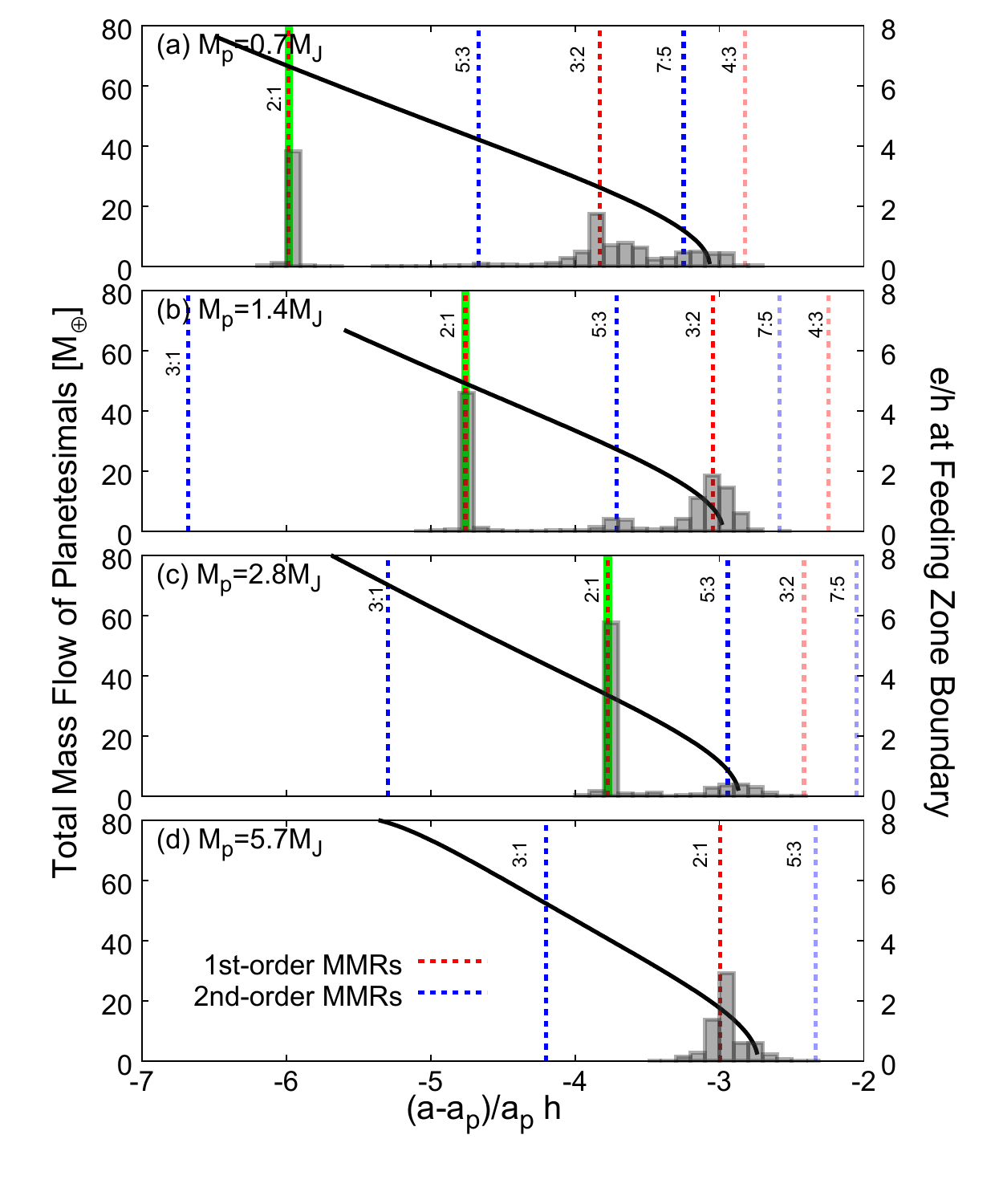}
        \caption{
            The total mass of planetesimals that enter the feeding zone (or the total mass flow of planetesimals) as a function of their location relative the planet at the time of entry, $\tilde{b}=(a-a_{\rm p})/a_{\rm p} h$ (see Eq.~[\ref{eq:reduced_Hill_radius}]) for the cases of the planet mass $M_{\rm p}$ = (a) $0.7M_{\rm J}$, (b) $1.4M_{\rm J}$, (c) $2.8M_{\rm J}$ and (d) $5.7M_{\rm J}$.
            Red dotted lines show the positions of 1st-order MMRs and blue dotted lines show the positions of 2nd-order MMRs. 
        The MMRs highlighted with the green line do not supply planetesimals into the feeding zone (see Appendix.~\ref{App_resonance}).
        }
        \label{fig:Result2_btilde}
        \end{center}
    \end{figure}

    Figure~\ref{fig:Result2_btilde} shows the total mass of planetesimals that enter the feeding zone vs.~$\tilde{b}$ at the time of entry.
    The positions of 1st-order MMRs such as 2:1 and 3:2 and 2nd-order MMRs such as 3:1 and 5:3 are indicated by the red and blue dotted lines, respectively. 
    The $\tilde{e}$-$\tilde{b}$ relationship for the feeding zone boundary is shown by the black solid line.

    Also, we highlight MMRs that do not contribute to the enrichment of the planet with heavy elements. 
    Planetesimals that enter the feeding zone via these MMRs are not captured by the planet. This is because the strongly trapped planetesimals leave the feeding zone through the same MMR, as shown in  Fig.~\ref{fig:Result1_Resonance_Shepherding2}. 

    First, we find that a large fraction of planetesimals enter the feeding zone through the 1st-order MMRs.
    Thus, in the cases of $M_{\rm p}=0.7~M_{\rm J}$ and $1.4~M_{\rm J}$, the 3:2 MMR is the main channel into the feeding zone.
    When the planetary mass is larger than $1.4~M_{\rm J}$, the 3:2 MMR is engulfed by the feeding zone and only a small amount of planetesimals enter the feeding zone through the 5:3 MMR, which results in the decrease in $\Mcaptot$. 
    Finally, in the case of $M_{\rm p}=5.7~M_{\rm J}$, the 2:1 MMR is located at $\tilde{b}~\sim~3$ and the equilibrium eccentricity becomes larger than the eccentricity corresponding to the feeding zone boundary. 
    As a result, the 2:1 MMR becomes the main channel supplying planetesimals into the feeding zone, and for the planetary enrichment. 

    Comparing the cases of $M_{\rm p}=0.7~M_{\rm J}$ and $1.4~M_{\rm J}$, we find that the total mass of planetesimals entering the feeding zone is larger in the former case ($\sim74~\Mearth$) than the latter case ($\sim66~\Mearth$); nevertheless, the total mass of captured planetesimals is larger in the latter case.
    As shown in Fig.~\ref{fig:Result2_btilde}, the 1st-order MMR (i.e. 3:2 MMR) is located at $\tilde{b}\sim-4$ for $M_{\rm p}=0.7~M_{\rm J}$ and $\tilde{b}\sim-3$ for $M_{\rm p}=1.4~M_{\rm J}$.
    An increase in $|\tilde{b}|$ corresponds to that in $\tilde{e}$ at the feeding zone boundary, which results in a decrease in the planetesimal capture probability.
    In summary, $\Mcaptot$ for $M_{\rm p}=1.4~M_{\rm J}$ is larger than that of $M_{\rm p}=0.7~M_{\rm J}$ because the 1st-order MMR (i.e. 3:2 MMR) is closer to the planet in the former case.

\begin{acknowledgements}
    We thank the referee for valuable comments.
    This work is supported by JSPS Core-to-Core Program "International Network of Planetary Sciences (Planet2)" and JSPS KAKENHI Grant Numbers 17H01153 and 18H05439.  
    R.H. acknowledges support from SNSF grant 200021\_169054.
    Some of this work has been carried out within the framework of the National Centre for Competence in Research PlanetS, supported by the Swiss National Foundation.
    Numerical computations were carried out on the Cray XC50 at the Center for Computational Astrophysics, National Astronomical Observatory of Japan.
\end{acknowledgements}


\bibliographystyle{aa}
\bibliography{refs}

\begin{thebibliography}{52}
\expandafter\ifx\csname natexlab\endcsname\relax\def\natexlab#1{#1}\fi

\bibitem[{{Adachi} {et~al.}(1976){Adachi}, {Hayashi}, \&
  {Nakazawa}}]{Adachi+1976}
{Adachi}, I., {Hayashi}, C., \& {Nakazawa}, K. 1976, Progress of Theoretical
  Physics, 56, 1756

\bibitem[{{Alibert} {et~al.}(2005){Alibert}, {Mordasini}, {Benz}, \&
  {Winisdoerffer}}]{Alibert+2005}
{Alibert}, Y., {Mordasini}, C., {Benz}, W., \& {Winisdoerffer}, C. 2005, \aap,
  434, 343

\bibitem[{{Alibert} {et~al.}(2018){Alibert}, {Venturini}, {Helled}, {Ataiee},
  {Burn}, {Senecal}, {Benz}, {Mayer}, {Mordasini}, {Quanz}, \&
  {Sch{\"o}nb{\"a}chler}}]{Alibert18}
{Alibert}, Y., {Venturini}, J., {Helled}, R., {et~al.} 2018, Nature Astronomy,
  2, 873

\bibitem[{Batygin \& Laughlin(2015)}]{Batygin+2015}
Batygin, K. \& Laughlin, G. 2015, Proceedings of the National Academy of
  Sciences of the United States of America, 112, 4214

\bibitem[{Bitsch {et~al.}(2019)Bitsch, Izidoro, Johansen, Raymond, Morbidelli,
  Lambrechts, \& Jacobson}]{Bitsch+2019a}
Bitsch, B., Izidoro, A., Johansen, A., {et~al.} 2019
  [\eprint[arXiv]{1902.08771v1}]

\bibitem[{{Booth} {et~al.}(2017){Booth}, {Clarke}, {Madhusudhan}, \&
  {Ilee}}]{Booth+2017}
{Booth}, R.~A., {Clarke}, C.~J., {Madhusudhan}, N., \& {Ilee}, J.~D. 2017,
  mnras, 469, 3994

\bibitem[{Brouwers {et~al.}(2018)Brouwers, Vazan, \& Ormel}]{Brouwers+2018}
Brouwers, M.~G., Vazan, A., \& Ormel, C.~W. 2018, Astrophysics A{\&}A, 611, 65

\bibitem[{Dullemond {et~al.}(2018)Dullemond, Birnstiel, Huang, Kurtovic,
  Andrews, Guzm{\'{a}}n, P{\'{e}}rez, Isella, Zhu, Benisty, Wilner, Bai,
  Carpenter, Zhang, \& Ricci}]{Dullemond+2018}
Dullemond, C.~P., Birnstiel, T., Huang, J., {et~al.} 2018, The Astrophysical
  Journal, 869, L46

\bibitem[{Goldreich \& Schlichting(2014)}]{Goldreich+2014}
Goldreich, P. \& Schlichting, H.~E. 2014, The Astronomical Journal, 147, 32

\bibitem[{{Guillot} {et~al.}(2006){Guillot}, {Santos}, {Pont}, {Iro}, {Melo},
  \& {Ribas}}]{Guillot+2006}
{Guillot}, T., {Santos}, N.~C., {Pont}, F., {et~al.} 2006, \aap, 453, L21

\bibitem[{Hands \& Alexander(2018)}]{Hands+2018}
Hands, T.~O. \& Alexander, R.~D. 2018, MNRAS, 474, 3998

\bibitem[{{Hasegawa} {et~al.}(2018){Hasegawa}, {Bryden}, {Ikoma}, {Vasisht}, \&
  {Swain}}]{Hasegawa+2018}
{Hasegawa}, Y., {Bryden}, G., {Ikoma}, M., {Vasisht}, G., \& {Swain}, M. 2018,
  \apj, 865, 32

\bibitem[{{Hayashi}(1981)}]{Hayashi1981}
{Hayashi}, C. 1981, Progress of Theoretical Physics Supplement, 70, 35

\bibitem[{{Hayashi} {et~al.}(1977){Hayashi}, {Nakazawa}, \&
  {Adachi}}]{Hayashi+1977}
{Hayashi}, C., {Nakazawa}, K., \& {Adachi}, I. 1977, \pasj, 29, 163

\bibitem[{{Helled} \& {Schubert}(2009)}]{Helled09}
{Helled}, R. \& {Schubert}, G. 2009, \apj, 697, 1256

\bibitem[{{Helled} \& {Stevenson}(2017)}]{HS2017}
{Helled}, R. \& {Stevenson}, D. 2017, \apjl, 840, L4

\bibitem[{{Hori} \& {Ikoma}(2011)}]{Hori+2011}
{Hori}, Y. \& {Ikoma}, M. 2011, \mnras, 416, 1419

\bibitem[{Humphries \& Nayakshin(2018)}]{Humphries+2018}
Humphries, R.~J. \& Nayakshin, S. 2018, MNRAS, 477, 593

\bibitem[{Ida \& Lin(2004)}]{Ida+2004a}
Ida, S. \& Lin, D. N.~C. 2004, The Astrophysical Journal, 604, 388

\bibitem[{Ida \& {Nakazawa}(1989)}]{Ida+1989}
Ida, S. \& {Nakazawa}, K. 1989, Astronomy {\&} Astrophysics, 2

\bibitem[{{Ikoma} {et~al.}(2006){Ikoma}, {Guillot}, {Genda}, {Tanigawa}, \&
  {Ida}}]{Ikoma+2006b}
{Ikoma}, M., {Guillot}, T., {Genda}, H., {Tanigawa}, T., \& {Ida}, S. 2006,
  \apj, 650, 1150

\bibitem[{{Ikoma} {et~al.}(2000){Ikoma}, {Nakazawa}, \& {Emori}}]{Ikoma+2000}
{Ikoma}, M., {Nakazawa}, K., \& {Emori}, H. 2000, apj, 537, 1013

\bibitem[{{Inaba} \& {Ikoma}(2003)}]{Inaba+2003}
{Inaba}, S. \& {Ikoma}, M. 2003, aap, 410, 711

\bibitem[{Johansen {et~al.}(2015)Johansen, Low, Lacerda, \&
  Bizzarro}]{Johansen+2015}
Johansen, A., Low, M. M.~M., Lacerda, P., \& Bizzarro, M. 2015, Science
  Advances, 1 [\eprint[arXiv]{1503.07347}]

\bibitem[{{Kanagawa} {et~al.}(2017){Kanagawa}, {Tanaka}, {Muto}, \&
  {Tanigawa}}]{Kanagawa+2017}
{Kanagawa}, K.~D., {Tanaka}, H., {Muto}, T., \& {Tanigawa}, T. 2017, pasj, 69,
  97

\bibitem[{Kokubo \& Ida(1998)}]{Kokubo+1998}
Kokubo, E. \& Ida, S. 1998, Icarus, 131, 171

\bibitem[{{Lambrechts} {et~al.}(2014){Lambrechts}, {Johansen}, \&
  {Morbidelli}}]{Lambrechts+2014}
{Lambrechts}, M., {Johansen}, A., \& {Morbidelli}, A. 2014, \aap, 572, A35

\bibitem[{Lissauer(1987)}]{Lissauer1987}
Lissauer, J.~J. 1987, Icarus, 69, 249

\bibitem[{{Liu} {et~al.}(2015){Liu}, {Agnor}, {Lin}, \& {Li}}]{Liu2015}
{Liu}, S.-F., {Agnor}, C.~B., {Lin}, D.~N.~C., \& {Li}, S.-L. 2015, \mnras,
  446, 1685

\bibitem[{{Lozovsky} {et~al.}(2017){Lozovsky}, {Helled}, {Rosenberg}, \&
  {Bodenheimer}}]{Lozovsky2017}
{Lozovsky}, M., {Helled}, R., {Rosenberg}, E.~D., \& {Bodenheimer}, P. 2017,
  \apj, 836, 227

\bibitem[{{Lynden-Bell} \& {Pringle}(1974)}]{Lynden-Bell+1974}
{Lynden-Bell}, D. \& {Pringle}, J.~E. 1974, \mnras, 168, 603

\bibitem[{{Malhotra}(1993)}]{Malhotra1993}
{Malhotra}, R. 1993, icarus, 106, 264

\bibitem[{{Miller} \& {Fortney}(2011)}]{Miller+2011}
{Miller}, N. \& {Fortney}, J.~J. 2011, \apjl, 736, L29

\bibitem[{Morbidelli {et~al.}(2009)Morbidelli, Bottke, Nesvorn{\'{y}}, \&
  Levison}]{Morbidelli+2009}
Morbidelli, A., Bottke, W.~F., Nesvorn{\'{y}}, D., \& Levison, H.~F. 2009,
  Icarus, 204, 558

\bibitem[{{Mordasini}(2014)}]{Mordasini2014}
{Mordasini}, C. 2014, \aap, 572, A118

\bibitem[{{Murray} \& {Dermott}(1999)}]{Murray+1999}
{Murray}, C.~D. \& {Dermott}, S.~F. 1999, {Solar system dynamics}

\bibitem[{Ormel(2013)}]{Ormel2013}
Ormel, C.~W. 2013, MNRAS, 428, 3526

\bibitem[{{Podolak} {et~al.}(1988){Podolak}, {Pollack}, \&
  {Reynolds}}]{Podolak+1988}
{Podolak}, M., {Pollack}, J.~B., \& {Reynolds}, R.~T. 1988, \icarus, 73, 163

\bibitem[{{Pollack} {et~al.}(1996){Pollack}, {Hubickyj}, {Bodenheimer},
  {Lissauer}, {Podolak}, \& {Greenzweig}}]{Pollack+1996}
{Pollack}, J.~B., {Hubickyj}, O., {Bodenheimer}, P., {et~al.} 1996, icarus,
  124, 62

\bibitem[{{Shakura} \& {Sunyaev}(1973)}]{Shakura+1973}
{Shakura}, N.~I. \& {Sunyaev}, R.~A. 1973, aap, 24, 337

\bibitem[{{Shibata} \& {Ikoma}(2019)}]{Shibata+2019}
{Shibata}, S. \& {Ikoma}, M. 2019, \mnras, 487, 4510

\bibitem[{{Shiraishi} \& {Ida}(2008)}]{Shiraishi+2008}
{Shiraishi}, M. \& {Ida}, S. 2008, apj, 684, 1416

\bibitem[{{Tanaka} \& {Ida}(1999)}]{Tanaka+1999}
{Tanaka}, H. \& {Ida}, S. 1999, icarus, 139, 350

\bibitem[{{Tanigawa} \& {Ikoma}(2007)}]{Tanigawa+2007}
{Tanigawa}, T. \& {Ikoma}, M. 2007, apj, 667, 557

\bibitem[{{Tanigawa} {et~al.}(2014){Tanigawa}, {Maruta}, \&
  {Machida}}]{Tanigawa+2014}
{Tanigawa}, T., {Maruta}, A., \& {Machida}, M.~N. 2014, apj, 784, 109

\bibitem[{{Tanigawa} \& {Tanaka}(2016)}]{Tanigawa+2016}
{Tanigawa}, T. \& {Tanaka}, H. 2016, \apj, 823, 48

\bibitem[{{Thorngren} {et~al.}(2016){Thorngren}, {Fortney}, {Murray-Clay}, \&
  {Lopez}}]{Thorngren+2016}
{Thorngren}, D.~P., {Fortney}, J.~J., {Murray-Clay}, R.~A., \& {Lopez}, E.~D.
  2016, apj, 831, 64

\bibitem[{{Valletta} \& {Helled}(2019)}]{Valletta2019}
{Valletta}, C. \& {Helled}, R. 2019, \apj, 871, 127

\bibitem[{{Venturini} {et~al.}(2015){Venturini}, {Alibert}, {Benz}, \&
  {Ikoma}}]{Venturini+2015}
{Venturini}, J., {Alibert}, Y., {Benz}, W., \& {Ikoma}, M. 2015, aap, 576, A114

\bibitem[{{W Beckwith} {et~al.}(1996){W Beckwith}, Sargent, \&
  ~}]{Beckwith+1996}
{W Beckwith}, S.~V., Sargent, A.~I., \& ~, S. 1996, {Circumstellar disks and
  the search for neighbouring planetary systems e. 0 4fi}, Tech. rep.

\bibitem[{Yu \& Tremaine(2001)}]{Yu+2001}
Yu, Q. \& Tremaine, S. 2001, The Astronomical Journal, 121, 1736

\bibitem[{{Zhou} \& {Lin}(2007)}]{Zhou+2007}
{Zhou}, J.-L. \& {Lin}, D.~N.~C. 2007, apj, 666, 447

\end{thebibliography}
\addcontentsline{toc}{section}{References}

\label{lastpage}

%
%

\end{document}